\documentclass[journal]{IEEEtran}

\usepackage{amsmath}
\usepackage{amssymb}
\usepackage{graphicx}
\usepackage{booktabs}
\usepackage{array}
\usepackage{url}
\usepackage[colorlinks=true,citecolor=blue,linkcolor=blue,urlcolor=blue]{hyperref}
\usepackage{orcidlink}   

\begin{document}

\title{\fontsize{19}{22}\selectfont Task-Based Evaluation of Raw Radar Data
Compression: A Pre-Registered Study of Where Classical Codecs Fail to
Preserve Target Detection, and~Why}

\author{Eric~Michael~Chrabot~\orcidlink{0009-0009-8363-7878}%
\thanks{E. M. Chrabot is with Signal Current Inc., Norfolk, VA, USA
(e-mail: eric@signalcurrent.ai).}%
\thanks{All work reported here was performed at private expense, on publicly
available data, prior to any government award. No classified or controlled
unclassified information was accessed or used at any point.}}

\markboth{Preprint, v3, November 2026}%
{Chrabot: Task-Based Evaluation of Raw Radar Data Compression}

\maketitle

\begin{abstract}
Synthetic aperture radar (SAR) systems, spaceborne and airborne, collect raw
in-phase/quadrature (I/Q) echo data at rates that exceed downlink and storage
capacity; fielded spaceborne systems compress onboard with block-adaptive
quantization (BAQ) and its flexible dynamic successor (FDBAQ). Proposed
replacements, including learned compression, are typically evaluated with
image-quality metrics such as peak signal-to-noise ratio (PSNR), structural
similarity (SSIM), or signal-to-quantization-noise ratio (SQNR), none of which
measures whether the data still supports its operational use. We introduce a
task-based evaluation methodology for raw radar compression, with a published
pre-registration trail, in which codecs are scored after SAR focusing against
a two-sided detection criterion (a constant false alarm rate (CFAR)
detection-retention floor and a spurious-detection budget at matched
threshold), using frozen task models trained once on uncompressed data. On a
Sentinel-1 stripmap Level-0 scene, which is FDBAQ output delivered at 5.22
bits per complex sample over the study crop, no tested configuration sustains
utility below that delivered rate: the best, 4-bit BAQ with entropy coding of its indices, needs
5.92 bits; the best transform codec on the registered grid, raw-domain HEVC,
7.62; and 3-bit BAQ at 6.12 bits fails the retention floor. On re-compressed
data no tested re-quantization is therefore certified below the delivered
rate, and the registered-grid frontier on this scene is robust to matching
tolerance, detector false-alarm rate, and added noise. Energy concentration after focusing delivers the predicted retention
gain at matched rate but, below about 6 bits, converts it into diffuse
spurious detections; an apparent concentration of them near strong scatterers
is traced to the study's plane mapping and withdrawn. The failure the second
clause exists to catch, a codec that passes the retention floor while flooding
the detector, appears on an Umbra 1\,m spotlight image (JPEG2000 at 2 bits per
complex sample retains 0.910 of reference detections while producing 15 times
the spurious budget); on AFRL Gotcha GMTI phase history, never compressed
onboard, the classical frontier is 8.0--10.9 bits per complex sample on a
259-detection reference. The complete pre-registration trail, every logged
configuration, and the evaluation harness are public, with corrections to the
record stated in place.
\end{abstract}


\section{Introduction}
\IEEEPARstart{N}{ext}-generation spaceborne and airborne radars, and
synthetic aperture radar (SAR) systems in particular, collect raw in-phase/quadrature (I/Q) echo
data at rates that routinely exceed downlink capacity. The fielded answer is
onboard block-adaptive quantization: BAQ, and its flexible-dynamic successor FDBAQ, which reduce raw echoes to roughly 2.6--3.6 bits per real component.
(Rates are reported in bits per complex sample throughout this paper, I and Q
combined, so the European Space Agency's (ESA) per-component figures for FDBAQ
correspond to roughly 5--7 bits per complex sample; the scene studied here was
delivered at 5.22.) BAQ has been the operational standard for over
three decades because it is simple, fast, and, as this paper independently
confirms, close to the rate-distortion bound for a raw echo source that is
well modeled sample-by-sample as memoryless complex Gaussian.

Learned compression is now an active research direction for this problem,
including on raw SAR data directly (Section~\ref{sec:related}). The question
these efforts raise is whether a learned codec can outperform BAQ/FDBAQ at
matched utility. Answering it requires deciding what ``utility'' means.
Nearly every published evaluation, for raw-SAR neural compression and
classical baselines alike, uses an image-quality proxy: peak signal-to-noise ratio (PSNR), structural similarity (SSIM), or signal-to-quantization-noise ratio (SQNR)
between reconstructed and reference imagery. These are convenient and easy to
report, but they do not measure what the data is collected to do. A
compression artifact that costs half a dB of SQNR at a scatterer is invisible
to a fidelity metric and can be catastrophic to a constant false alarm rate (CFAR) detector reading the same pixel; conversely, an artifact that a
fidelity metric penalizes heavily may leave detection performance untouched.
The Air Force Research Laboratory (AFRL) made this point about classical SAR compression in 2010: standard
image-quality metrics are ``much less stringent'' than a metric tied to
coherent change detection performance (Scarborough et al.,
2010~\cite{r6}). This paper gives that observation a concrete,
pre-registered, falsifiable form.

\textit{Contribution~1: a task-based evaluation methodology.} We define
``sustains utility at rate $R$'' as a two-sided criterion, scored after SAR
focusing rather than on raw or range-compressed echoes: a codec must retain
at least 90\% of a CFAR detector's detections on the uncompressed reference
(written $P_d$ below for continuity with the record; it is a retention rate
against the reference detector, not an absolute detection probability against
surveyed targets), and must hold spurious (unmatched) detections to at most
10\% of the reference count, at identical CFAR configuration. Both conditions
are scored against frozen task models trained once on uncompressed data
before any codec sweep runs. The retention floor, the physics prediction
motivating post-focus scoring, and the interpretation rules for each
subsequent result were committed before the corresponding results were read;
the spurious-detection clause was committed minutes later, after the first
sweep's rows had been written but before any was read.
Section~\ref{sec:prereg} gives the chronology; we mark the clause an amendment
declared before analysis rather than a blind pre-registration wherever it is
used.

\textit{Contribution~2: an empirically characterized classical frontier on
re-compressed and on unencoded data.} Applied to a Sentinel-1 stripmap scene
whose Level-0 input is itself FDBAQ output at 5.22 bits per complex sample
over the study crop, no tested codec sustains utility below that delivered
rate: entropy-coded 4-bit BAQ needs 5.92 bits, raw-domain HEVC on the
registered grid 7.62, and 3-bit BAQ fails the floor (Sections~\ref{sec:harness}
and~\ref{sec:frontier}). That is the correct answer for a re-compression
scenario, and it is what the previous versions of this paper mis-stated as a
frontier well below the fielded point. Raw-domain scalar quantization has no
transform-coding gain to exploit (Section~\ref{sec:concentration}); transform
coding in the focused domain exploits that gain but, below about 6 bits per
complex sample, pays for it in diffuse spurious detections, and above it
sustains without beating the entropy-coded scalar quantizer
(Section~\ref{sec:posthoc}); a
partial transform (range dechirp without azimuth focusing) is insufficient on
its own (Section~\ref{sec:ablation}). On unencoded airborne data
(Section~\ref{sec:gotcha}) the frontier is 8.0--10.9 bits per complex sample,
held by raw-domain HEVC.

\textit{Contribution~3: a validity check against a known artifact.} The
closest prior work (Asiyabi et al., 2025~\cite{r1}) reports that on
FDBAQ-decoded Sentinel-1 data, 3-bit BAQ can score anomalously well because
the FDBAQ reconstruction lattice in the data happens to align with the 3-bit
BAQ step. The artifact is tested for directly with and without their
lattice-adaptation procedure: retention changes by at most 0.007 at every bit depth, spurious counts by 0.0\% at the 3-bit point and by 2--12\% elsewhere, bit-depth monotonicity holds with and without the adaptation, and 3-bit BAQ fails the floor under both (Section~\ref{sec:harness}).

\textit{Contribution~4: a bounded negative result for a small learned codec},
reported under interpretation rules fixed before the model was trained
(Section~\ref{sec:learned}), together with a statement of the open problem the
classical results imply: capture the coding gain that domain concentration
provides without inheriting the artifact signature that destroys detection.

\textit{Scope.} This is a feasibility study centered on one pre-registered
scene plus two further scenes, an airborne replication and an image-domain
spotlight scene, not a benchmark suite (Section~\ref{sec:limitations}). We consider the methodological record
(pre-registration with a documented amendment and retraction trail, and a
stopping rule that ended the exploratory phase once the classical baselines
and one trained model had been scored) to be as much a part of the
contribution as the empirical results; both are reported in full rather than
as a cleaned-up narrative.

\section{Related Work}
\label{sec:related}
\textit{Learned compression of raw SAR data.} The closest prior work is
Asiyabi et al. (2025)~\cite{r1}, the current state of the art for learned
compression of raw SAR data. They train a fully complex-valued convolutional
autoencoder (four CV convolutional layers with generalized divisive
normalization, CV-GDN) with a factorized entropy model, quantizer, and
arithmetic coder, directly on raw Sentinel-1 Level-0 data, and evaluate
reconstructions after SAR focusing using SQNR, phase error, and complex
coherence between focused single-look-complex (SLC) products. Their workflow
(compress raw data, decompress, focus to SLC, then measure quality on the
focused product) independently arrived at the same post-focus scoring
discipline adopted here, and is taken as external support for that choice,
which was reached after first scoring on unfocused data and finding the
result dominated by speckle rather than detection behavior
(Section~\ref{sec:postfocus}). They report substantial bitrate reductions at
matched signal fidelity (roughly 42\% fewer bits than BAQ at 15~dB SQNR, 27\%
at 20~dB) and demonstrate cross-sensor generalization, training on Sentinel-1
and testing on ERS-1. The present study differs from theirs in two ways.
First, their metrics (SQNR, phase error, complex coherence) are
signal-fidelity measures; their evaluation contains no detector, false-alarm
count, or downstream task model. Second, they identify but do not resolve a
training/evaluation mismatch in their setup: ``distortion loss is computed
between the input and output SAR raw data, as the SLC products are not
available while training the network.'' They train on raw-domain mean squared error (MSE) and
evaluate in the focused domain; the two-sided detection criterion used here,
and the open problem it exposes (Section~\ref{sec:learned}), bear directly on
that gap. Their FDBAQ adaptation procedure is also adopted, from their
companion IGARSS 2024 paper~\cite{r2}: adding uniform noise to fill the gaps a
prior FDBAQ pass leaves in the reconstruction lattice, which restores
statistics closer to unquantized raw data for evaluation purposes. Because
the harness-validation claim rests on the 3-bit BAQ configuration, the
reconstruction-lattice artifact they report at that bit depth is tested for
directly and found to change retention by at most 0.007 at any bit depth and
spurious counts by nothing at the 3-bit point (Section~\ref{sec:harness}).

\textit{Classical transform-domain compression of raw SAR.} Frequency-domain
BAQ (FFT-BAQ) applies block-adaptive quantization after an FFT rather than in
the time domain, recovering roughly 1 bit per sample over time-domain BAQ at
matched fidelity (Fischer, Benz, and Moreira, DLR, IGARSS 1999)~\cite{r3}.
Onboard systems nonetheless continued to fly time-domain BAQ variants, with
implementation complexity the commonly cited reason. A 2025 classical
baseline applies channel-adaptive serial BAQ in the range-Doppler domain---a
classical quantizer operating in nearly the same transformed domain as the
dechirp-only ablation arm of Section~\ref{sec:ablation} (Jiang et al.,
2025)~\cite{r4}. Both are direct ancestors of the ``compress after a domain
transform'' idea used here, and the historical complexity objection to
FFT-BAQ is precisely the deployability question the compute-ablation and
operation-count accounting of Section~\ref{sec:deploy} is designed to answer
for a learned codec.

\textit{Task- and change-detection-aware SAR compression evaluation.} Novak
and Frost (2009) measure the effect of SAR image compression directly on
coherent change detection (CCD) performance rather than on image
fidelity~\cite{r5}; in study design this is the closest ancestor of the
present work. AFRL's 2010 challenge-problem paper makes the same argument
explicitly: classical compression schemes ``generally use visual image
quality as the metric of performance, which is much less stringent than a
metric relating to quality SAR CCD,'' and it poses ``develop methods to
compress the complex SAR images while maintaining quality SAR CCD'' as an open
challenge problem with a public dataset (Scarborough et al., 2010)~\cite{r6}.
Task-based compression evaluation for SAR is therefore an established framing
within AFRL itself, sixteen years before this study. The same paper's
reference list also indexes the classical SAR compression literature this
work builds against: entropy-constrained quantization of raw SAR
data~\cite{r18}, complex-valued SAR image compression~\cite{r19}, and
tree-structured wavelet coding~\cite{r20}.

\textit{Classical phase-history compression.} Trellis-coded quantization of
SAR phase history~\cite{r7} and wavelet-packet compression of phase-history
data (Pascazio and Schirinzi, IGARSS 2000)~\cite{r8} establish that classical
compression of pre-image-formation radar data is a 1990s-era research line,
not a gap. We are not aware of a \textit{learned} compression method
operating on phase-history or raw echo data that has been evaluated against a
detection or tracking task, which is the gap this paper addresses
empirically. The same appears true one representation downstream: a
literature search conducted for this work found learned methods applied to
range-profile recognition but not to range-profile compression.

\textit{Task-aware compression, more broadly, and its information-theoretic
lineage.} Task-aware compression for natural images and for SAR imagery
specifically is an active area (target-perception-aware SAR image
compression~\cite{r9}; neural feature compression for satellite
downlink~\cite{r10}; onboard semantic compression~\cite{r11}; and, for
complex-valued RF communication signals, learned compression evaluated by
downstream modulation-classification accuracy~\cite{r22}). We do not claim to
have invented the idea of compressing for a downstream task rather than for
reconstruction fidelity. The idea has a specific information-theoretic
lineage: indirect (remote) rate-distortion theory~\cite{r12}, the information
bottleneck~\cite{r13}, hardware-limited task-based quantization~\cite{r14},
and information-distilling quantizers, designed to preserve mutual information
about a downstream inference target rather than reconstruction
fidelity~\cite{r15}. The nearest radar-side member of this lineage is BiLiMO
(Xi, Shlezinger, and Eldar, 2021)~\cite{r21}, which applies task-based
quantization at the acquisition stage of a MIMO radar receiver, designing
bit-limited ADC front ends from which target parameters remain recoverable.
To our knowledge this lineage has not previously been applied to compression
of recorded raw SAR data for storage or downlink, nor evaluated against a
post-focus detection criterion. Framing the two-sided criterion within this
sixty-year line of work situates it as an instance of a general principle,
compressing for the inference the data will be used for, rather than an ad hoc
metric invented for one study.

\textit{Scope of the claimed contribution.} Each ingredient above has an
ancestor. What we did not find in this literature is (a) a two-sided
detection criterion, combining a detection-retention floor and a
spurious-detection budget at matched CFAR configuration, used as a compression
acceptance test (existing task-aware SAR work uses CCD quality or
single-sided detection metrics); (b) such a criterion applied with an
auditable pre-registration trail, including amendments and a documented
retraction; (c) application to raw echo-domain data with post-focus scoring,
rather than to already-focused imagery; or (d) a learned codec evaluated
under such a criterion together with an explicit parallel-entropy-coding
deployability constraint. The claim here is to this combination, with each
ingredient's ancestors cited above, not to any single ingredient in
isolation.

\section{Methodology}

\subsection{Problem formulation and utility criterion (declared before analysis)}
\label{sec:formulation}
\textit{Notation.} Let $\mathbf{x} \in \mathbb{C}^{M \times N}$ denote a block
of raw complex echo data ($M$ pulses by $N$ range samples). A codec at rate
$R$ is a map $\hat{\mathbf{x}} = \mathcal{C}_R(\mathbf{x})$; the rate is
measured in bits per complex sample,
\begin{equation}
R = \frac{B}{MN}, \qquad \rho = \frac{64}{R},
\label{eq:rate}
\end{equation}
where $B$ is the total encoded bit count and $\rho$ is the compression ratio
against the native representation (two 32-bit floating-point components per
complex sample). Let $\mathcal{F}$ be the SAR focusing operator, so
$\mathbf{I} = \mathcal{F}(\mathbf{x})$ is the focused reference image and
$\hat{\mathbf{I}} = \mathcal{F}(\hat{\mathbf{x}})$ the focused reconstruction.
All utility is scored on $\hat{\mathbf{I}}$, not on $\hat{\mathbf{x}}$
(Section~\ref{sec:postfocus}).

\textit{Detector.} On a power image $P = |\mathbf{I}|^2$, a two-dimensional
cell-averaging CFAR estimates local clutter power $\mu_{ij}$ as the mean of
$P$ over a training ring (an outer window of side $2(g+t)+1$ minus an inner
guard window of side $2g+1$, for $g$ guard and $t$ training cells per side),
and declares a detection where
\begin{equation}
D_{ij} = \mathbb{1}\!\left[\,P_{ij} > \alpha\,\mu_{ij}\,\right],
\qquad
\alpha = n_t\!\left(P_{\mathrm{fa}}^{-1/n_t} - 1\right),
\label{eq:cfar}
\end{equation}
with $n_t$ the number of training cells and $P_{\mathrm{fa}}$ the design
false-alarm probability; $\alpha$ is the standard cell-averaging CFAR (CA-CFAR) threshold factor for
exponentially distributed clutter power. Reducing $D$ to local-maximum
locations yields a detection point set $\mathcal{D}(\mathbf{I})$.

\textit{Detection agreement.} Writing $\mathcal{D}_{\mathrm{ref}} =
\mathcal{D}(\mathbf{I})$ and $\mathcal{D}_{\mathrm{rec}} =
\mathcal{D}(\hat{\mathbf{I}})$, and using a spatial match tolerance $\tau$,
the retention rate (written $P_d$) and spurious count at rate $R$ are
\begin{equation}
P_d(R) = \frac{\bigl|\{\, p \in \mathcal{D}_{\mathrm{ref}} :
\min_{q \in \mathcal{D}_{\mathrm{rec}}} \lVert p - q \rVert \le \tau \,\}\bigr|}
{\bigl|\mathcal{D}_{\mathrm{ref}}\bigr|},
\label{eq:pd}
\end{equation}
\begin{equation}
N_{\mathrm{spur}}(R) = \bigl|\{\, q \in \mathcal{D}_{\mathrm{rec}} :
\min_{p \in \mathcal{D}_{\mathrm{ref}}} \lVert q - p \rVert > \tau \,\}\bigr|.
\label{eq:spur}
\end{equation}
$P_d$ measures retention of the reference detector's output, and
$N_{\mathrm{spur}}$ counts reconstruction detections with no reference
counterpart. Detections are reduced to $3\times3$ local maxima before
matching; a reference detection is retained if any reconstruction detection
lies within Euclidean distance $\tau$ (in cells) of it, and a reconstruction
detection is spurious if no reference detection lies within $\tau$; the norm
in \eqref{eq:pd} and \eqref{eq:spur} is $\ell_2$. Matching is
nearest-neighbour, not one-to-one: several reconstruction peaks near one
reference peak count as one retention and zero spurious detections, which is
lenient toward the flooding failure mode reported below. The reference set
contains, by construction, the detector's own false detections on
uncompressed speckle: at the design $P_{\mathrm{fa}}$ of $10^{-4}$ over the
22.0 million scored interior cells of the Illinois scene the expected number
of threshold exceedances on homogeneous clutter is about 2{,}200 pixels before
the local-maximum reduction, about 16\% of the 13{,}964 reference peaks under
that design assumption (heterogeneous clutter raises the CA-CFAR false-alarm
rate above design, and the $P_{\mathrm{fa}}$ sweep of Section~\ref{sec:posthoc}
implies a larger threshold-set share); the same sweep shows that the frontier
and its sustaining set do not change as that share changes. The reference set is
the detector's own output on the uncompressed data and is not validated
against surveyed targets (Section~\ref{sec:limitations}).

\textit{Two-sided utility criterion.} A codec \textit{sustains utility at rate
$R$} if and only if
\begin{equation}
P_d(R) \ge P_d^{\min}
\quad\text{and}\quad
N_{\mathrm{spur}}(R) \le \beta\,\bigl|\mathcal{D}_{\mathrm{ref}}\bigr|,
\label{eq:crit}
\end{equation}
with $(P_d^{\min}, \beta) = (0.9,\, 0.1)$ and the detector configuration
$(P_{\mathrm{fa}}, g, t, \tau) = (10^{-4},\, 2,\, 8,\, 3)$ held identical
across every codec. The thresholds $(0.9, 0.1)$ are round numbers chosen
without reference to any result: the 0.9 floor was committed with the first
sweep in progress and the 0.1 budget after that sweep had finished but before
any row was read (Section~\ref{sec:prereg}). The rationale is one of
symmetry and convention rather than optimization. A retention floor of 0.9
is the customary high-retention bar for a detection acceptance test; the
spurious budget of 10\% of the reference count bounds the growth of the
detector's output, and so of the downstream load, at the same relative
tolerance the floor allows for loss, so that the two clauses admit the same
fraction of error in each direction. The detector configuration was fixed
once on the uncompressed reference, before any codec sweep, and never
adjusted per codec: a design $P_{\mathrm{fa}}$ of $10^{-4}$ is a conventional
value; a 2-cell guard excludes the point-target mainlobe from the training
ring; 8 training cells per side give $n_t = 416$, for which the threshold
factor $\alpha = 9.31$ is within 1.1\% of its large-sample limit
$-\ln P_{\mathrm{fa}} = 9.21$; and a match tolerance of 3 cells is three
times the focuser's measured $-3$~dB width of 1.0 cell (Section~\ref{sec:data}),
chosen before the width was measured, to absorb sub-cell peak shifts;
Section~\ref{sec:posthoc} shows every frontier unchanged at 1, 2, and 5 cells. None of these values was tuned
against any result reported here. Section~\ref{sec:frontier} reports the frontier's
sensitivity to the two thresholds; Section~\ref{sec:posthoc} reports, post
hoc, its sensitivity to the match tolerance, to the design
$P_{\mathrm{fa}}$, and to added noise. The single-sided alternative---thresholding on $P_d$
alone---is shown in Sections~\ref{sec:concentration} and~\ref{sec:umbra} to
certify two image-domain configurations producing 1.7 and 15 times the
tolerable spurious count; the
second condition in \eqref{eq:crit} is what rejects them. The retention floor was committed before any result was read;
the spurious-detection clause was committed after the first sweep's rows had
been written (Section~\ref{sec:prereg}).

\textit{Classical frontier and GO threshold.} Over the classical codecs
evaluated,
\begin{equation}
R_c = \min_{\mathcal{C}\in\text{classical}}
\{\, R : \mathcal{C} \text{ sustains utility at } R \,\},
\quad
R_{\mathrm{GO}} = \tfrac{1}{2} R_c.
\label{eq:rc}
\end{equation}
The pre-registered GO condition for a learned codec is to sustain utility at a
rate no greater than $R_{\mathrm{GO}}$, i.e.\ at least a $2\times$ rate
advantage over the best classical configuration. Because the rule is
rate-relative, it is fixed independently of where $R_c$ falls.

\begin{figure*}[t]
\centering
\includegraphics[width=0.92\textwidth]{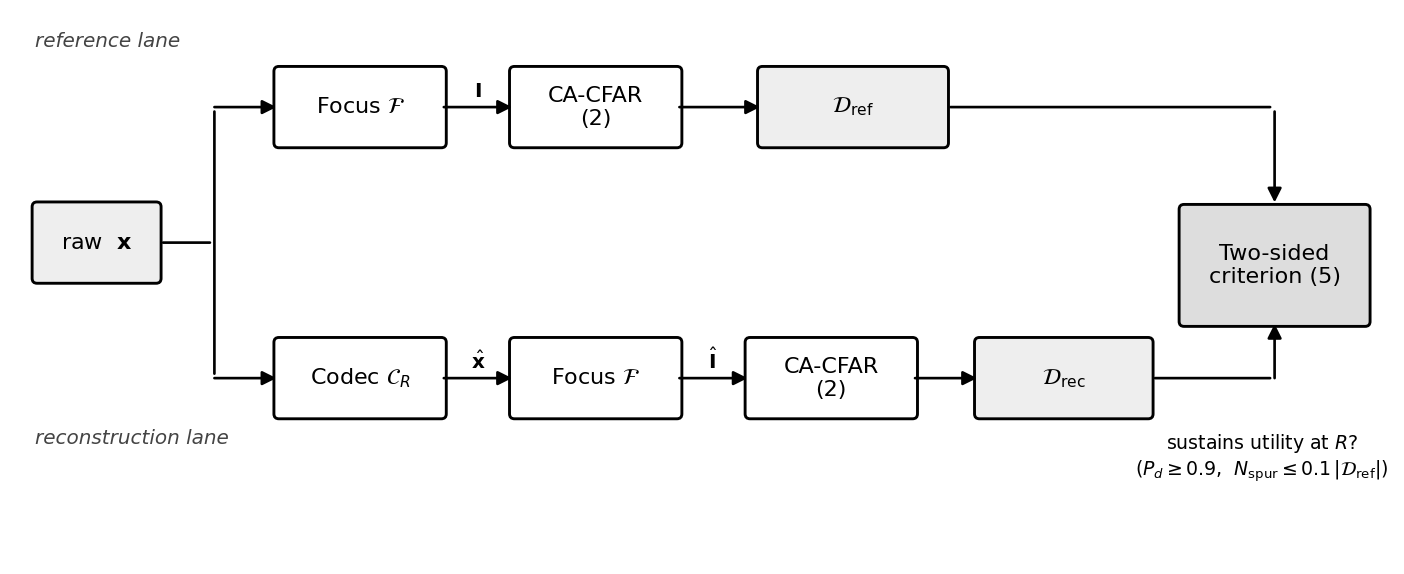}
\caption{The evaluation pipeline. A raw echo block $\mathbf{x}$ is carried
down two lanes. The reference lane focuses it ($\mathcal{F}$) and detects
(CA-CFAR, \eqref{eq:cfar}), yielding $\mathcal{D}_{\mathrm{ref}}$. The
reconstruction lane first compresses and decompresses ($\mathcal{C}_R$), then
focuses and detects the same way, yielding $\mathcal{D}_{\mathrm{rec}}$. The
two detection sets feed the two-sided criterion \eqref{eq:crit}, which
determines whether the codec sustains utility at rate $R$. Detection is scored
on the focused images $\mathbf{I}, \hat{\mathbf{I}}$, never on the raw or
reconstructed echoes (Section~\ref{sec:postfocus}).}
\label{fig:pipeline}
\end{figure*}

\subsection{Why post-focus scoring}
\label{sec:postfocus}
CFAR applied to unfocused or merely range-compressed echo data thresholds
speckle rather than signal; an early version of this harness scored detection
that way, and the resulting numbers were effectively noise. The reason to
score after focusing is physical. SAR focusing is coherent integration with
substantial processing gain: approximately white quantization noise
integrates incoherently while target returns integrate coherently, so
focusing should substantially suppress the effect of compression noise on
detection. This prediction was recorded before the corrected sweep ran
(Section~\ref{sec:prereg}) and held. The original pre-focus numbers, obtained
on a different Sentinel-1 scene (interferometric-wide mode, Hampton Roads) and
retracted as an operational-utility claim (Section~\ref{sec:limitations}),
gave retention of 0.2--0.4 at about 4 bits per complex sample, against
0.76--0.84 post-focus on the Illinois crop at 4.1--4.8 bits; the two numbers
come from different scenes and modes and are not a controlled comparison.

\subsection{Frozen task models}
\textit{Detection.} CA-CFAR as specified in Section~\ref{sec:formulation},
applied to the focused image.

\textit{Recognition (image-domain complement).} A small convolutional
classifier for automatic target recognition (ATR): three stride-2
convolutional layers (1$\to$32, 32$\to$64, 64$\to$128 channels; $5\times5$,
$5\times5$, $3\times3$ kernels) with ReLU, global pooling, and a linear
128$\to$3 head, trained once on uncompressed MSTAR chips ($17^\circ$
depression, three classes: BMP2, BTR70, T72) and evaluated, frozen, on 300 of
the 1{,}365 held-out $15^\circ$-depression chips against every codec operating
point (sampling standard error about 2 percentage points). As a robustness
check against the objection that a weak classifier would show exaggerated
sensitivity to compression artifacts, a strengthened classifier was trained
(shift augmentation, 60 epochs, cosine learning-rate schedule; uncompressed
accuracy on the 300 chips 86.0\% $\rightarrow$ 89.3\%) and the full sweep
rerun: the degradation curve's shape is unchanged (e.g., High Efficiency Video
Coding (HEVC) at 1.38 bits per complex sample: 47.7\% $\rightarrow$ 48.0\%;
JPEG2000 at 0.98 bits per complex sample: 55.3\% $\rightarrow$ 57.7\%). All
accuracies reported in Section~\ref{sec:atr} and Figure~\ref{fig:atr} are from
the strengthened classifier.

\subsection{Pre-registration protocol}
\label{sec:prereg}
Pre-registration, a practice adopted from experimental science, means
committing a study's endpoints, predictions, and analysis rules to an
immutable, timestamped record before the corresponding data exist, so that
apparent success cannot be manufactured after the fact by selecting the
metric, threshold, or data subset that happens to look best. The risk is
concrete for a compression study: given the freedom to choose the rate, the
detector configuration, and the pass/fail line after seeing results, almost
any codec can be made to look good. Here, endpoints, physics predictions, and
interpretation rules are committed, with timestamps, before the data they
concern exists, and every amendment and retraction is recorded in place
rather than edited out. We distinguish
\textit{declared-before-analysis} from \textit{provably blind}, and the
distinction applies to the criterion itself. The first focused-domain sweep
ran as a background process that appended a row to the results log as each
operating point completed. The commit declaring the retention floor
(\texttt{08c420d}, 00:41:58 local time) was made with six of the twelve
first-sweep rows already on disk; the commit adding the spurious-detection
budget, the second clause of \eqref{eq:crit} (\texttt{22df565}, 00:44:52),
was made with all twelve on disk, under a commit message that wrongly stated
none existed. That statement is retracted in the record. What the timestamps
establish is what was on disk at each commit; that no row was read before
either commit (the results were first opened after 00:45) rests on the
author's account, recorded in the notebook, and cannot be verified from the
history. The
retention floor is therefore declared before analysis with partial data on
disk, and the spurious-detection clause, which this paper's central finding
rests on, is an amendment declared before analysis, not a blind
pre-registration. A skeptical reader should weight both accordingly, and we
mark the clause's status wherever it is used. Later amendments, such as the
model-training interpretation rules of Section~\ref{sec:learned}, were fixed
before any training checkpoint existed and so achieve blindness by
construction rather than by assertion. We consider this documented
imperfection more informative than a narrative that omits it.

\subsection{Rate convention}
All rates in this paper are reported in bits per complex sample, per the
definition of $R$ in \eqref{eq:rate}, and compression ratios $\rho$ against
the native 64-bit complex representation. The convention matters because
ESA's FDBAQ documentation quotes rates per real component (roughly 2.6--3.6
bits per component in interferometric-wide mode), which is 5--7 bits per
complex sample on this convention; the stripmap scene studied here was
delivered at 5.22 bits per complex sample over the crop and 5.28 over the full
range line, consistent with a low-backscatter agricultural scene coded at the
lowest bit-rate code throughout the crop (Section~\ref{sec:harness}). Any
comparison to a per-component figure must double it first.

\section{Data, Transform, and Baselines}
\label{sec:data}
\textit{Three datasets, three roles.} This paper uses three radar datasets
with distinct roles, and every result is labeled with its scene
(Table~\ref{tab:datasets}). The three $R_c$ values are not comparable to each
other, since sensor, geometry, scene, data domain, and statistical weight all
differ (Sections~\ref{sec:gotcha} and~\ref{sec:umbra}), and no claim in this
paper rests on comparing them. What carries across datasets is the
qualitative structure: a classical frontier exists below the rates a fidelity
metric would suggest, and energy concentration trades retention against a
spurious-detection cost, on all three.

\begin{table*}[t]
\centering
\caption{The three datasets and their roles.}
\label{tab:datasets}
\renewcommand{\arraystretch}{1.25}
\setlength{\tabcolsep}{3pt}
\begin{tabular}{@{}p{0.155\textwidth} p{0.265\textwidth} p{0.265\textwidth} p{0.245\textwidth}@{}}
\toprule
 & \textbf{Sentinel-1 (Illinois)} & \textbf{AFRL Gotcha GMTI} & \textbf{Umbra (Port of Vancouver)} \\
\midrule
Role & Primary study, declared before analysis (Sec.~\ref{sec:harness}--\ref{sec:learned}) & Replication of the concentration/spurious-detection trade, raw domain (Sec.~\ref{sec:gotcha}) & Image-domain extension, exploratory (Sec.~\ref{sec:umbra}) \\
Platform, domain & Spaceborne, C-band, stripmap; raw echoes, focused before scoring & Airborne, X-band, motion-compensated phase history; range-Doppler image formed before scoring & Spaceborne, X-band, 1\,m spotlight; delivered complex image (SICD), no focusing step \\
Encoding history & FDBAQ applied onboard (re-compression scenario) & Never compressed onboard & Vendor-processed product; onboard handling not disclosed \\
Provenance & Civil/ESA open data & Government-provided (AFRL public release) & Umbra open data (CC BY 4.0) \\
Reference detections (peaks) & 13{,}964 & 259 & 22{,}860 \\
Fielded-codec anchor & Delivered FDBAQ bitstream 5.22 bps over the crop, below every sustaining configuration (Sec.~\ref{sec:harness}) & None exists for this sensor & None; BAQ fails at every tested rate \\
Classical frontier $R_c$ & 7.62 bits/complex-sample on the registered grid (raw HEVC); 5.92 with entropy-coded 4-bit BAQ & 8.0--10.9 bits/complex-sample (bracket; reference-composition dependent) & 3.87 bits/complex-sample (HEVC, under the clipped mapping) \\
\bottomrule
\end{tabular}
\end{table*}

\textit{Data.} The primary scene is a Sentinel-1D S6 stripmap raw acquisition
over north-central Illinois (2026-08-05; frame footprint
$40.64^\circ$--$42.37^\circ$N, $88.97^\circ$--$87.68^\circ$W per the SAFE
manifest---agricultural and exurban terrain west of Chicago), decoded from
Level-0 packets and cropped to a $4096\times8192$ raw region containing
13{,}964 reference detections (CFAR exceedance peaks) under the frozen CFAR
configuration on the uncompressed, focused reference. Sentinel-1 Level-0 data has already passed
through onboard FDBAQ before distribution, so every result on this scene
measures re-compression of operationally compressed raw data. That is the
actual spaceborne downlink scenario, and it is the scope of the claim. The
caveat is specific to downlink-limited spaceborne data and does not extend to
public raw SAR in general: AFRL's Gotcha ground moving target indication (GMTI) phase history shows no such
quantization structure (153{,}443 distinct sample values out of 153{,}600 in
a representative block, i.e.\ effectively unquantized), and
Section~\ref{sec:gotcha} evaluates it directly. The third scene is a
$4096\times4096$ centre crop of the public Umbra Port of Vancouver spotlight
product (Umbra-07, 2025-12-10, HH, 1\,m), a delivered complex image in
Sensor-Independent Complex Data (SICD) format with 22{,}860 reference
detections (vessels, port infrastructure, and urban structure over dark
water); codecs are applied directly to the complex pixels under the same plane
mapping and CFAR configuration, with no focusing step
(Section~\ref{sec:umbra}). It was evaluated on 2026-08-06, after the stopping
rule, and is reported as exploratory.

\begin{figure*}[t]
\centering
\includegraphics[width=0.98\textwidth]{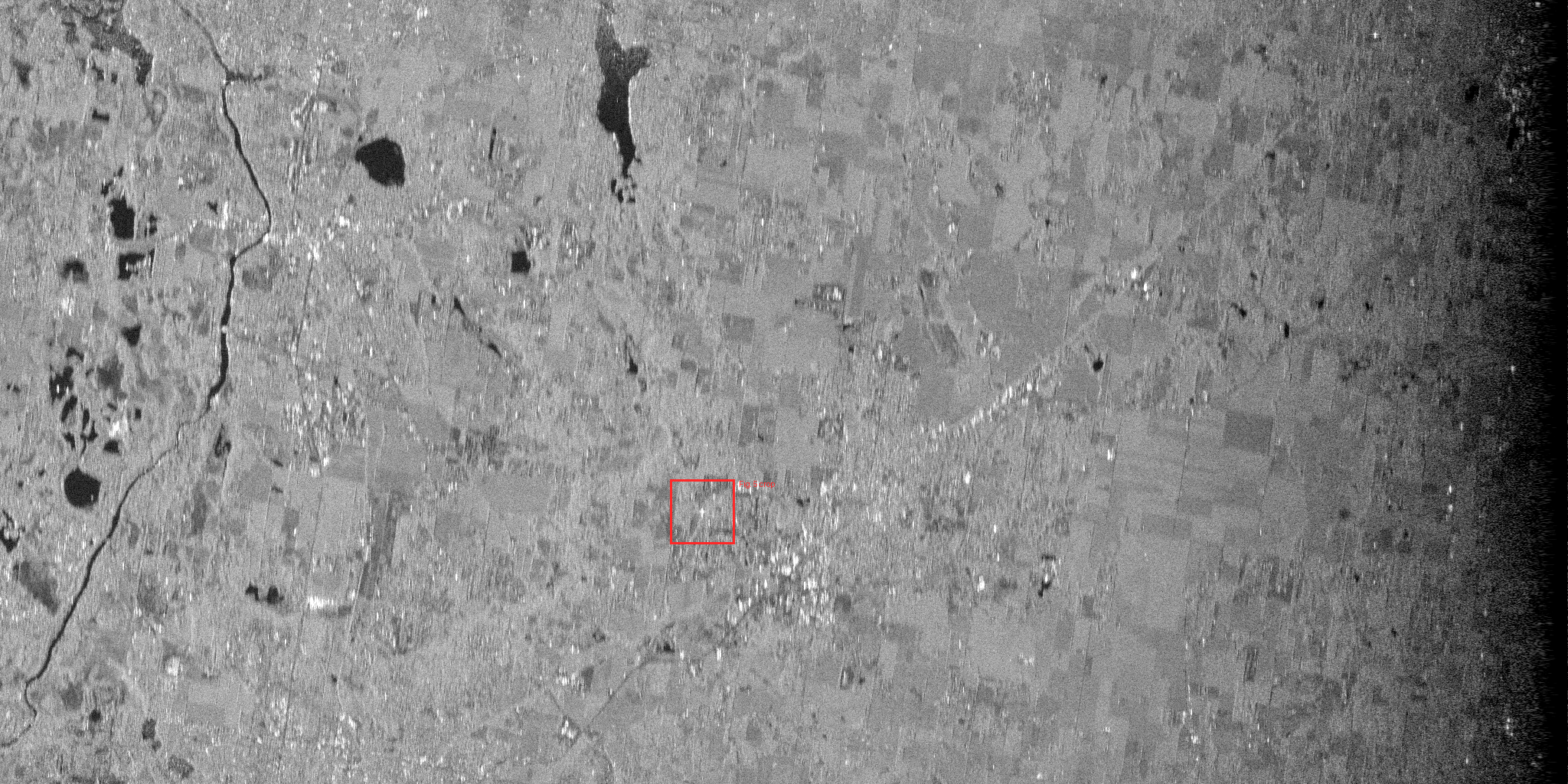}
\caption{The focused reference scene (north-central Illinois stripmap,
log-intensity quicklook) that every rate, $P_d$, and spurious-detection number
for the primary study is computed against. Predominantly agricultural
terrain---fields (mid-gray checkerboard texture), scattered ponds and a river
(dark), and sparse rural road/settlement structure (bright linear
features)---with no dense urban core or large lake in this crop. Single
clutter regime, per the limitation stated in
Section~\ref{sec:limitations}. Red box: the Figure~\ref{fig:ringing} crop
($320\times320$ cells centred at row 2{,}672, column 3{,}668 of the focused
image).}
\label{fig:scene}
\end{figure*}

\textit{Focuser.} A range-Doppler-algorithm focuser is built directly from
packet-header chirp and timing parameters (transmit pulse rate/length,
range-decimation sample rate). The Doppler centroid is taken as zero. The focuser is validated visually
and by the impulse-response measurement of Section~\ref{sec:posthoc}; no
comparison against the agency's Level-1 product of the same frame has been
made. The result of Section~\ref{sec:harness}, that no tested codec sustains
the criterion below the delivered bitstream rate, is an empirical finding and
is not offered as a validation of the focuser.

\textit{Unitary dechirp/focus transform.} Raw echoes are near-memoryless
sample to sample and therefore give scalar quantization little to improve on
(Section~\ref{sec:concentration}). To give any codec cross-sample structure to
exploit, an exactly invertible transform is used: a phase-only matched filter,
linear-phase range-cell-migration correction, and a phase-only azimuth filter,
applied as circular convolutions so that every stage is unitary by
construction. Round-trip invertibility (forward transform, then inverse, on
uncompressed data) was verified numerically before anything was built on top
of it, at a mean relative error below $10^{-6}$ for complex64 (the test
tolerance is $5\times10^{-6}$; the Gotcha transform of
Section~\ref{sec:gotcha} measured $4.4\times10^{-7}$). Verification matters here because
the naive matched filter $\mathrm{conj}(\mathrm{FFT}(\text{replica}))$ has
magnitude collapse outside the chirp band and is not invertible; the
phase-only variant $H/|H|$ is required to pass the round-trip test. The
transform's role is confined to the codec: it defines the domain in which
the transform-domain arms compress, and the exact inverse is applied before
scoring, so every detection number on the Sentinel-1 scene, for every arm,
is computed on the output of the range-Doppler focuser above after the exact
inverse, never on the transform domain itself (on the Gotcha scene the
transform domain coincides with the scored image domain,
Section~\ref{sec:gotcha}; the Umbra scene has no focusing step). Its focusing
quality therefore does not enter any score. Measured post hoc on a unit point
target synthesised at the crop centre by the transform inverse and focused by
the range-Doppler focuser, the point-spread function has a $-3$~dB width of
1.0 cell in both range and azimuth and peak sidelobe ratios of $-14.9$~dB
(range) and $-13.3$~dB (azimuth), an unweighted sinc-class response confined
to a few cells (Section~\ref{sec:posthoc}).

\textit{Baselines.} Lloyd--Max (Gaussian-optimal, non-uniform) BAQ at
1/2/3/4/6 bits; JPEG2000 and HEVC applied to the I and Q planes independently
under a single, fixed, fairness-audited mapping (symmetric percentile clip at
the 99.99th percentile of $|\text{value}|$ per plane, clip bounds carried as
side information, then uint16) applied identically to every plane-based codec
in every domain---raw and transformed. This mapping was not a neutral default:
an earlier min/max mapping produced degenerate near-black HEVC frames on this
heavy-tailed data, and the percentile-clip fix shifts effective distortion
toward uniform relative error, which more closely matches how a CFAR threshold
behaves---a change that could materially favor HEVC, so it is applied
identically to every codec and stated explicitly rather than left implicit.
BAQ, as the fielded reference, operates natively on floating-point samples
with no such mapping and takes no benefit from it. Every plane-based baseline
is also evaluated inside the unitary transform, in both a dechirp-only and a
full-focus configuration (Section~\ref{sec:ablation}).

\section{Results}

\subsection{Harness validation, and a test against a known confound}
\label{sec:harness}
The Level-0 input is FDBAQ output. Its own bitstream, parsed block by block from the
user data of the 4{,}096 echoes that make up the crop, costs 5.22 bits per
complex sample over the crop's 8{,}192 range samples (sample codes plus block
side information; 5.19--5.26 across echoes; every block inside the crop
carries the lowest bit-rate code, and the full 20{,}072-sample range line,
whose far-range blocks use higher codes, averages 5.28), and it reconstructs
the input exactly:
retention 1.0, zero spurious detections, by identity. That row heads
Table~\ref{tab:illinois}, and every sustaining configuration sits above it
(every tested configuration below it fails the floor). Four-bit BAQ
sustains utility at 8.12 bits per complex sample ($P_d$ 0.937, 799 spurious
detections against a budget of 1{,}396.4), or at 5.92 bits with its quantizer
indices entropy coded (identical reconstruction, Section~\ref{sec:posthoc});
3-bit BAQ does not (6.12 bits fixed-rate, 5.34 entropy coded, $P_d$ 0.842). A
memoryless lossless coder of the input's 80-level alphabet would cost 10.18
bits. The reading is the one a re-compression scenario makes plausible, and it is
an empirical result rather than a property the criterion guarantees: a lossy
re-quantizer could in principle sustain the criterion below the stream's own
rate, and none tested does, by 0.7 bits for the best entropy-coded scalar
quantizer and by 2.4 bits for the best transform codec on the registered
grid. In the previous versions of this paper the pass line was
reported between 2-bit and 3-bit BAQ, below the fielded point, because of the
scorer error described in the version note; the claim that the harness
``brackets the fielded operating point'' is withdrawn.

Because the comparison depends on BAQ at 3 and 4 bits specifically, a
confound reported by the closest prior work is tested directly (Asiyabi et
al., 2025~\cite{r1}; Section~\ref{sec:related}): on FDBAQ-decoded Sentinel-1
data, 3-bit BAQ can score anomalously well, even better than 4-bit BAQ,
because the FDBAQ reconstruction lattice already present in the data
coincidentally aligns with the BAQ quantization step at 3 bits. The
precondition is confirmed present in the data used here (48 distinct I values
across a 1{,}048{,}576-sample block and 80 across the full crop; 32 distinct
values in a representative $64\times64$ block, versus 153{,}443 out of
153{,}600 for AFRL's unencoded Gotcha data, Section~\ref{sec:data}); the BAQ
sweep was then rerun with and without the Asiyabi et al.\ lattice-adaptation
procedure (uniform-noise gap filling, verified to raise the distinct-value
count in a $64\times64$ block from 32 to 4{,}096), under the corrected scorer.
The result is unaffected at the points that matter: maximum $|\Delta P_d|$ is
0.007 across all four tested bit depths (3-bit BAQ 0.842 $\to$ 0.849, still
below the floor; 4-bit 0.937 $\to$ 0.940, still above it); spurious counts
move by $+1.8\%$, $0.0\%$, $+4.4\%$, and $+11.7\%$ at 2, 3, 4, and 6 bits; and
monotonicity in bit depth is preserved under both conditions. These eight
rows are in the results log (\texttt{study = rescore\_lattice}). The
3-bit-beats-4-bit inversion that is the artifact's signature is not observed.
Two properties of the setup likely explain this: the utility metric is
post-focus CFAR detection agreement, and coherent integration suppresses fine
quantization-lattice structure that a direct fidelity metric would register
immediately; and the BAQ implementation is Lloyd--Max (non-uniform), so its
reconstruction levels do not coincide with a uniform FDBAQ step.

\subsection{The classical frontier}
\label{sec:frontier}
Three frontiers are stated for this scene, because the corrected record
supports three different statements and a reader needs all of them. On the
registered codec grid (raw-domain scalar quantization at fixed rate;
plane-based transform coding in the raw and focused domains; dechirp-only and
full-focus variants), the best-performing configuration is raw-domain HEVC at
QP~28, which sustains utility at $R_c = 7.62$ bits per complex sample ($P_d$
0.945, 739 spurious detections), with raw-domain JPEG2000 at ratio 4 (8.00
bits, $P_d$ 0.938) and 4-bit BAQ (8.12 bits, $P_d$ 0.937) beside it; the next
points down, 3-bit BAQ (6.12 bits, $P_d$ 0.842) and HEVC at QP~36 (4.75 bits,
$P_d$ 0.839), fail the retention floor. This registered frontier is not
threshold-marginal: its six-member sustaining set is identical at matching
tolerances of 1, 2, 3, and 5 cells, at design $P_{\mathrm{fa}}$ of $10^{-3}$,
$10^{-4}$, and $10^{-5}$, and under added noise at every level tested
(Section~\ref{sec:posthoc}), and it is threshold-dependent in the expected
direction (Table~\ref{tab:thresholds}). Second, a post hoc focused-domain HEVC
arm at QP~20 sustains at 6.56 bits ($P_d$ 0.914, 1{,}219 spurious) under the
study's plane mapping, but only at the registered setting, at a tolerance
of 5 cells, and at $P_{\mathrm{fa}} = 10^{-5}$; it fails at tolerances of 1 and
2 cells and at $P_{\mathrm{fa}} = 10^{-3}$, so it is reported as the tested
minimum of the
transform codecs and not as a robust frontier. Third, entropy coding the
indices of the 4-bit BAQ row, at identical reconstruction and therefore
identical scores, lowers its rate from 8.12 to 5.92 bits
(Section~\ref{sec:posthoc}); that row's scores are the registered row's by
construction, its rate is an order-0 entropy bound rather than a measured
coder output, and it is the lowest sustaining rate of any tested codec on this
scene, 0.70 bits above the delivered FDBAQ bitstream. The bound is specific to
re-compression of the delivered lattice: under added noise the same row's
index entropy rises to 7.69 bits at the noisiest level tested
(Section~\ref{sec:posthoc}). The index entropy is low because the
input is itself FDBAQ output on an 80-level lattice; on unencoded Gotcha data
entropy coding gains only 0.1--0.4 bits (Section~\ref{sec:posthoc}). $R_c$ is a
property of this scene, sensor, processing lineage (FDBAQ-decoded Sentinel-1
stripmap), mapping rule, and tested codec set, not a universal constant.
Section~\ref{sec:gotcha} finds a different, higher frontier on unencoded
airborne data (AFRL Gotcha), but that comparison is confounded by sensor,
geometry, and scene differences in addition to encoding history; the
controlled version, $R_c$ on unencoded Gotcha data against the same data after
simulated FDBAQ, remains untested (Section~\ref{sec:conclusion}). Under the
pre-registered rate-relative rule, the registered-grid $R_c$ fixes the GO
threshold for a learned codec at $R_c/2 = 3.81$ bits per complex sample; a
literal application of the minimum in \eqref{eq:rc} to the extended codec set
gives $R_c = 5.92$ and $R_c/2 = 2.96$, and the threshold is kept at the grid
the rule was declared on. Either threshold lies below the delivered bitstream
(5.22), so a learned codec meeting it on this scene would have to beat the
fielded coder outright.
Table~\ref{tab:illinois} lists the operating points.

\begin{table}[tb]
\centering
\caption{Sentinel-1 (Illinois): selected classical operating points under the corrected scorer (peak reduction) and the corrected HEVC pipeline; the full set of 31 rows is in the results log. Rate in bits per complex sample; SQNR is image-domain, dB; the spurious budget is 1396.4 (10\% of 13{,}964 reference detections). ``Sust.'' = both clauses of \eqref{eq:crit} hold.}
\label{tab:illinois}
\renewcommand{\arraystretch}{1.15}
\setlength{\tabcolsep}{2pt}
\footnotesize
\resizebox{\columnwidth}{!}{%
\begin{tabular}{@{}llrrrrc@{}}
\toprule
Codec & Domain & Rate & SQNR & $P_d$ & Spur. & Sust. \\
\midrule
FDBAQ bitstream (as delivered, crop) & raw & 5.22 & $\infty$ & 1.000 & 0 & yes \\
Lossless (memoryless entropy) & raw & 10.18 & $\infty$ & 1.000 & 0 & yes \\
HEVC (QP12) & raw & 14.13 & 39.2 & 0.991 & 118 & yes \\
BAQ (6-bit) & raw & 12.12 & 30.7 & 0.977 & 282 & yes \\
HEVC (QP20) & raw & 10.85 & 31.1 & 0.979 & 294 & yes \\
BAQ (4-bit) & raw & 8.12 & 21.6 & 0.937 & 799 & yes \\
JPEG2000 (r4) & raw & 8.00 & 21.8 & 0.938 & 759 & yes \\
HEVC (QP28) & raw & 7.62 & 22.6 & 0.945 & 739 & yes \\
BAQ (3-bit) & raw & 6.12 & 14.1 & 0.842 & 1{,}635 & no \\
HEVC (QP36) & raw & 4.75 & 13.9 & 0.839 & 1{,}701 & no \\
JPEG2000 (r8) & dechirp & 4.00 & 10.3 & 0.761 & 2{,}671 & no \\
JPEG2000 (r8) & focused & 4.00 & 10.4 & 0.858 & 5{,}219 & no \\
JPEG2000 (r16) & focused & 2.00 & 5.4 & 0.820 & 16{,}546 & no \\
HEVC (QP36) & focused & 1.00 & 3.5 & 0.629 & 11{,}510 & no \\
JPEG2000 (r32) & focused & 1.00 & 3.2 & 0.775 & 31{,}142 & no \\
JPEG2000 (r64) & focused & 0.50 & 1.8 & 0.768 & 74{,}611 & no \\
JPEG2000 (r64) & raw & 0.50 & 1.4 & 0.252 & 3{,}795 & no \\
Learned (best) & focused & 0.89 & -- & 0.251 & 9{,}825 & no \\
BAQ (6-bit) + entropy coding & raw & 8.15 & 30.7 & 0.977 & 282 & yes \\
BAQ (4-bit) + entropy coding & raw & 5.92 & 21.6 & 0.937 & 799 & yes \\
BAQ (3-bit) + entropy coding & raw & 5.34 & 14.1 & 0.842 & 1{,}635 & no \\
\bottomrule
\end{tabular}}
\end{table}

\begin{table}[tb]
\centering
\caption{Sensitivity of the classical frontier to the criterion's thresholds, corrected scorer, computed from the logged rows (no new runs): lowest sustaining rate (bits per complex sample) and the codec that holds it (B$n$: $n$-bit BAQ; J$r$: raw JPEG2000 ratio $r$; H$q$: raw HEVC QP $q$; Jf/Hf: focused domain); $\beta$ is the spurious budget as a fraction of reference detections. Bold: the thresholds used in this paper.}
\label{tab:thresholds}
\renewcommand{\arraystretch}{1.15}
\setlength{\tabcolsep}{2pt}
\scriptsize
\resizebox{\columnwidth}{!}{%
\begin{tabular}{@{}lccc@{}}
\toprule
 & $\beta = 0.05$ & $\beta = 0.10$ & $\beta = 0.20$ \\
\midrule
\multicolumn{4}{@{}l}{\textit{Sentinel-1}} \\ \\
$P_d^{\min} = 0.85$ & 10.85 (H20) & 7.62 (H28) & 7.62 (H28) \\
$P_d^{\min} = 0.90$ & 10.85 (H20) & \textbf{7.62 (H28)} & 7.62 (H28) \\
$P_d^{\min} = 0.95$ & 10.85 (H20) & 10.85 (H20) & 10.85 (H20) \\
\multicolumn{4}{@{}l}{\textit{AFRL Gotcha GMTI}} \\ \\
$P_d^{\min} = 0.85$ & 10.93 (H12) & 7.95 (H20) & 7.95 (H20) \\
$P_d^{\min} = 0.90$ & 10.93 (H12) & \textbf{10.93 (H12)} & 7.99 (Jf4) \\
$P_d^{\min} = 0.95$ & 10.93 (H12) & 10.93 (H12) & 10.93 (H12) \\
\bottomrule
\end{tabular}}
\end{table}

\begin{figure}[tb]
\centering
\includegraphics[width=\columnwidth]{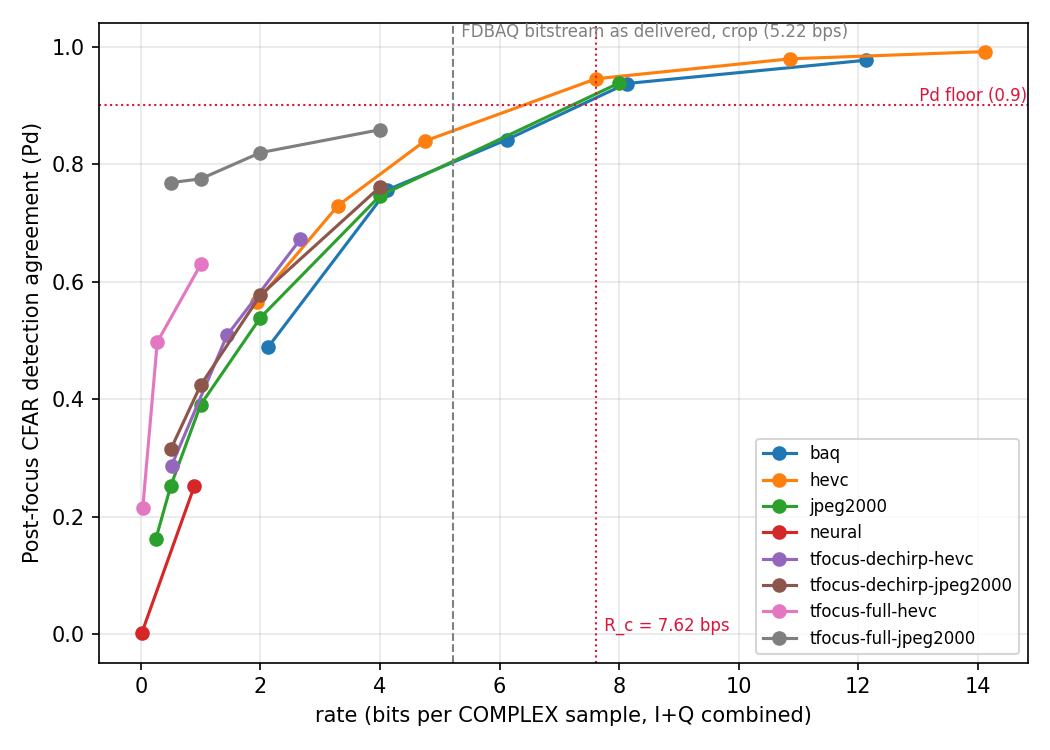}
\caption{Post-focus CFAR detection agreement ($P_d$) vs.\ rate, all classical
configurations and the trained learned codec. Horizontal dotted line: the
$P_d$ floor (0.9) of \eqref{eq:crit}. Vertical dotted line: $R_c = 7.62$ bits per
complex sample (raw-domain HEVC, the classical frontier). Corrected scorer. Vertical dashed gray
line: the delivered FDBAQ bitstream of this crop (5.22 bits per complex
sample).}
\label{fig:pd}
\end{figure}

\subsection{The concentration result and its spurious-detection cost}
\label{sec:concentration}
Coding-gain theory for unitary transforms is standard. A unitary transform
preserves total entropy but redistributes variance across coefficients, and
for a subsequent scalar quantizer the available transform-coding gain over $K$
coefficients with variances $\sigma_k^2$ is the ratio of their arithmetic to
geometric mean,
\begin{equation}
G_{\mathrm{TC}} =
\frac{\frac{1}{K}\sum_{k=1}^{K}\sigma_k^2}
{\left(\prod_{k=1}^{K}\sigma_k^2\right)^{1/K}}
\;\ge\; 1,
\label{eq:gain}
\end{equation}
with equality if and only if the variance spectrum is flat. Raw echo samples
are, sample by sample, near-memoryless complex Gaussian: their variance
spectrum is nearly flat, so $G_{\mathrm{TC}} \approx 1$ and there is
essentially no coding gain to exploit. This is why BAQ, a plain scalar
quantizer, is close to optimal in the raw domain, and why FDBAQ exists and
works. The focused-image domain is the opposite case: energy concentrates into
sparse bright scatterers over dark clutter, giving a large variance disparity,
$G_{\mathrm{TC}} \gg 1$, and in principle substantial coding gain. The gain
was not computed on the data; the argument is the standard one.

The data confirm the first half of the theory directly: full-focus JPEG2000
holds a matched-rate retention advantage over raw-domain JPEG2000, with $P_d$
0.858/0.820/0.775 versus 0.746/0.539/0.390 at 4/2/1 bits per complex sample.
Concentration does help retention at matched rate, as the coding-gain argument
predicts.

\begin{figure}[tb]
\centering
\includegraphics[width=\columnwidth]{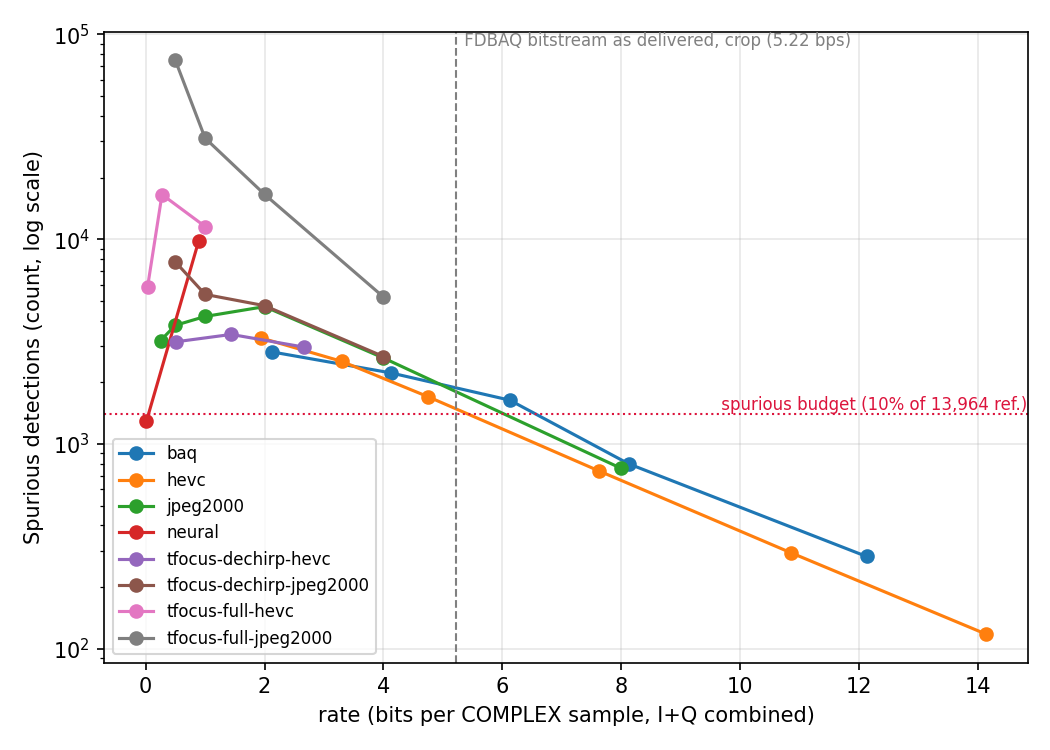}
\caption{Spurious (unmatched) CFAR detections vs.\ rate, log scale, all
classical configurations. Dotted red line: the 1{,}396.4-detection budget
(10\% of 13{,}964 reference detections). Every configuration that concentrates
energy in the focused domain sits well above budget at low rates (3.7 to 53
times for JPEG2000, 4 to 12 times for HEVC)---the same regime where
Figure~\ref{fig:pd} shows those configurations winning on $P_d$. Corrected
scorer.}
\label{fig:spur}
\end{figure}

The data also show a second effect. Focused SAR imagery consists of bright
point scatterers over dark speckle, and transform coding of that structure
produces distortion that a CFAR detector reads as new detections. Real
targets are largely retained ($P_d$ 0.768--0.858 across the tested rates)
while spurious counts grow sharply as rate falls: 5{,}219, 16{,}546,
31{,}142, and 74{,}611 spurious detections at 4, 2, 1, and 0.5 bits per
complex sample, against a budget of 1{,}396.4 (10\% of the 13{,}964 reference
detections). On this scene the retention floor rejects every one of these
points as well, so the two clauses of \eqref{eq:crit} agree on every
configuration; in the previous versions of this paper the 4-bit point
retained 0.912 under the pixel-count scorer and was the certified-and-flooding
example, and the corrected count puts it at 0.858. The certification failure
the second clause exists to catch is demonstrated on the Umbra scene
(Section~\ref{sec:umbra}). Figure~\ref{fig:spur} shows the shape of the
effect: every configuration that concentrates energy sits far above the
spurious-detection budget at low rates, while the same configurations are
the strongest performers on the $P_d$ axis of Figure~\ref{fig:pd} at those
rates. Neither figure alone shows why these codecs fail.

\begin{figure*}[t]
\centering
\includegraphics[width=0.92\textwidth]{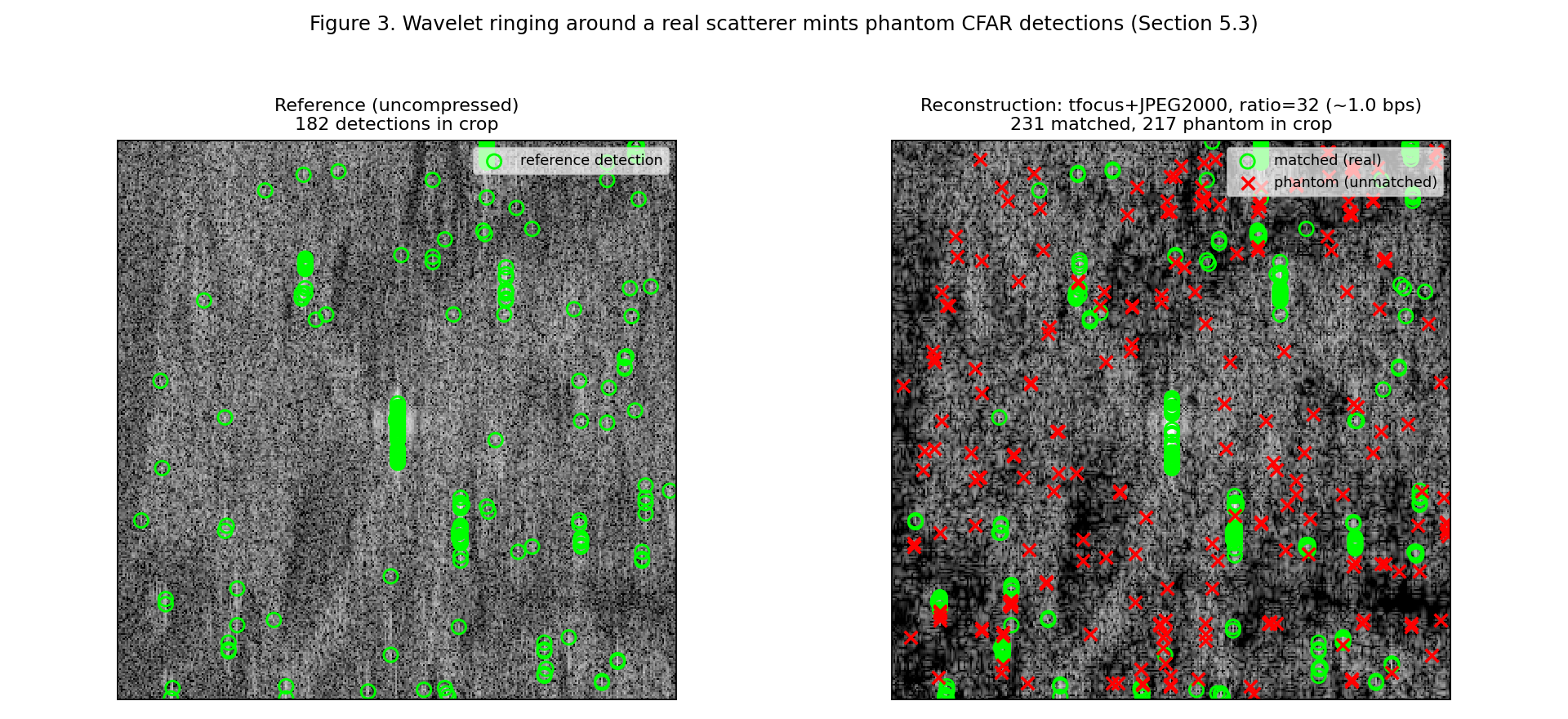}
\caption{A $320\times320$-pixel crop at the focused-domain JPEG2000, ratio-32
operating point (measured rate 1.000 bps; under the corrected scorer the full
region yields 31{,}142 phantom peaks against 13{,}964 reference peaks). Left: the uncompressed reference, with CFAR detections on real
scatterers circled in green. Right: the reconstruction after compress,
decompress, and focus. The same real scatterers are still detected (green),
but the surrounding neighborhood is now dense with phantom detections (red
$\times$) that do not exist in the reference. The neighbourhood was selected
for its phantom density and shows a neighbourhood of a strong scatterer; the
ablation of Section~\ref{sec:posthoc} finds no reliable excess of wavelet
phantoms near strong scatterers, and most lie far from any. The crop is
marked on Figure~\ref{fig:scene}.}
\label{fig:ringing}
\end{figure*}

The effect is not specific to the wavelet codec. Focused-domain HEVC, a
block-transform codec with no wavelet stage, produces 11{,}510, 16{,}520, and
5{,}847 spurious detections (4--12 times the budget) at the three rates of the
registered sweep (1.00, 0.27, and 0.03 bits), and the same pattern recurs for
HEVC on the Gotcha and Umbra scenes (Sections~\ref{sec:gotcha}
and~\ref{sec:umbra}). Where the spurious detections lie differs between the
two codecs, and the first version of this paper guessed wrong about it. A
post hoc ablation (Section~\ref{sec:posthoc}, Table~\ref{tab:ringing}) counts
the phantoms within 8 cells of the 1{,}397 strongest reference peaks against
the same count around the 1{,}397 weakest and around 1{,}397 random cells. For
the wavelet codec at 4 bits the three counts are 143, 72, and 62 of 5{,}219
phantoms, an excess near strong scatterers that falls short of the two-to-one
margin the test was declared with and that reverses at 2 bits (171 near
strong against 255 near weak); for HEVC they are 117, 9, and 10 at 8 bits.
Both excesses are the plane mapping's, not the codecs': the mapping's
no-codec round trip through the transform domain produces 184 phantoms of
which 113 lie near strong scatterers and one near weak ones
(Section~\ref{sec:posthoc}), because its clip saturates exactly the strongest
scatterers, and every transform arm carries that same 110--180 near-strong
population at every rate while its diffuse population scales with rate. Net of
the mapping, neither codec concentrates its spurious detections near strong
scatterers; the wavelet's are diffuse at every rate, HEVC's are few and
diffuse. Replacing the 9/7 filter with the reversible 5/3 filter changes
nothing. Whether the diffuse phantoms arise from clutter cells rising over the
CA-CFAR threshold or from the training-ring estimate beneath them falling was
not measured. Figure~\ref{fig:ringing} shows one dense neighbourhood at the
wavelet's 1-bit point, chosen for its phantom density, and is not
representative of where most spurious detections lie.

The effect, stated compactly: concentrating energy makes compression
distortion land across the image where a detector reads it as targets. The
retention floor alone catches it on this scene; the spurious-detection clause
is what catches it where retention survives (Section~\ref{sec:umbra}).

\subsection{Compute ablation: is full azimuth focusing necessary?}
\label{sec:ablation}
Full onboard SAR focusing is computationally the more expensive half of the
processing chain; if a low-complexity partial transform captured most of the
available coding gain, that would be the preferred configuration for embedded
deployment. A dechirp-only arm is tested (one complex multiply per sample plus
an FFT, collapsing the transmitted chirp to a peak per scatterer, without
azimuth focusing) against the full-focus transform. Full focus outperforms
dechirp-only at every matched rate tested ($P_d$ 0.820 versus 0.577 at 2 bits
per complex sample; the retention gap is 0.10, 0.24, 0.35, and 0.45 at 4, 2,
1, and 0.5 bits per complex sample). A pre-registered rate-equivalence test
was written to check whether the two arms could reach matched sustained
utility at comparable rates, but neither arm sustains utility at any rate in
the registered sweep (post hoc, focused-domain HEVC does sustain from 6.6
bits per complex sample, Section~\ref{sec:posthoc}, outside the grid this
test was declared on), so that test could not be evaluated as written. Under
the documented surrogate substituted for it, comparing $P_d$ at matched rate,
the gap between full focus and dechirp-only is outside the roughly 10\% band
the original test would have required at every rate (11\% at 4 bits, 30--59\%
at 2 bits and below). The conclusion holds under the surrogate: on this scene,
the partial transform is insufficient, and the coding gain that matters
requires full azimuth focusing.

\subsection{Image-domain complement: ATR under compression}
\label{sec:atr}
Focused-image classification is considerably more compression-tolerant than
raw-echo detection. With the strengthened classifier (ceiling 89.3\% on 300
chips, standard error about 2 points), JPEG2000 at 3.9 bits per complex
sample holds 89.7\% and HEVC at 3.5 bits holds 88.3\%, both within sampling
error of the ceiling. Collapse begins below roughly 2 bits per complex sample
(HEVC at 1.4: 48.0\%; JPEG2000 at 1.0: 57.7\%). BAQ, strong in the raw domain
precisely because it is matched to a near-memoryless source, is the weakest
performer on already-focused imagery below about 6 bits (58.3\% at 4.1 bits
per complex sample, 31 points below the ceiling; 82.7\% at 6.1 bits), while at
8 bits and above all three codecs are within about two standard errors of the
ceiling (BAQ 85.3\% at 8.1 bits, four points below it). This is the mirror image of the raw-domain result,
and consistent with the same coding-gain argument: BAQ has nothing to exploit
in a concentrated domain either, but there it is competing against codecs that
do.

\begin{figure}[tb]
\centering
\includegraphics[width=\columnwidth]{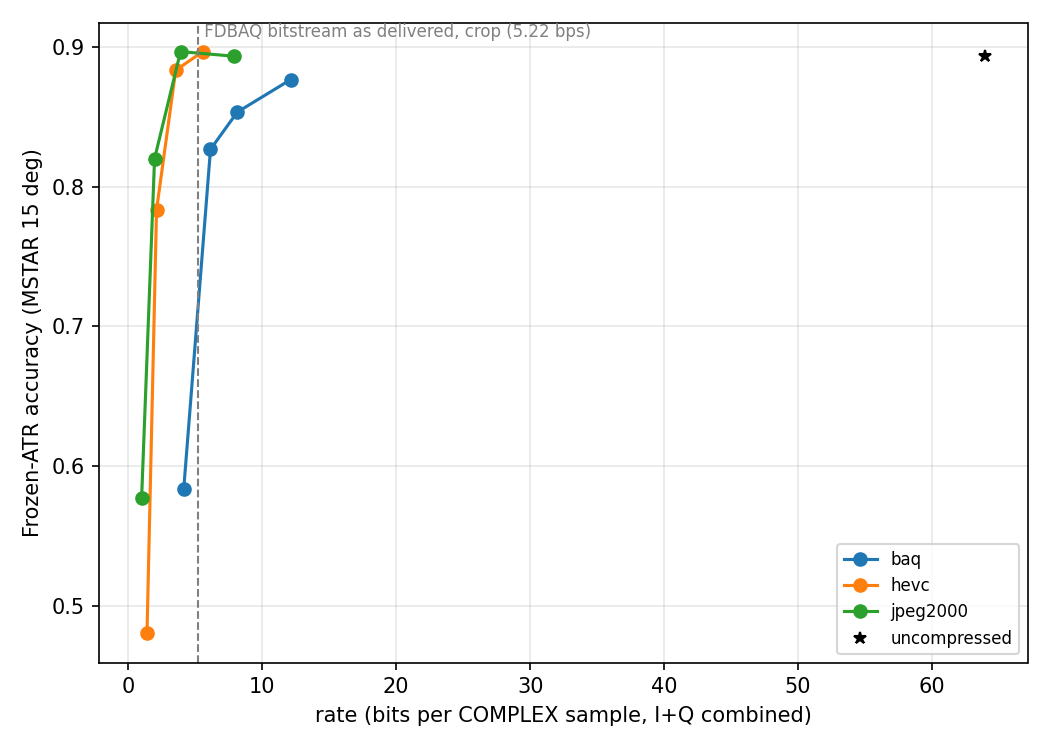}
\caption{Frozen-ATR accuracy (strengthened classifier, MSTAR, $15^\circ$
depression, 300 evaluation chips) vs.\ rate. BAQ trails JPEG2000 and HEVC
below about 6 bits per complex sample and is the weakest codec at 4 bits; at 8
bits and above all codecs are within about two standard errors of the
uncompressed ceiling (black star)---the mirror image of Figures~\ref{fig:pd}
and~\ref{fig:spur}, where BAQ is the strongest classical performer.}
\label{fig:atr}
\end{figure}

\subsection{Deployability}
\label{sec:deploy}
Measured wall-clock timings carry a named confound: the BAQ implementation
runs in-process as vectorized NumPy, while JPEG2000 and HEVC run through
external binaries with process-spawn and file I/O overhead, which understates
their relative cost. Implementation-independent operation-count classes are
therefore also reported, which are what actually transfers to embedded
hardware costing: BAQ, roughly $10$ operations/sample (normalize, a handful of
comparisons, scale), plus a few more for an entropy coder of its indices;
JPEG2000, roughly $10^2$--$10^3$ (wavelet cascade plus bitplane coding); HEVC,
roughly $10^2$--$10^3$ or more, and serial by construction, since its CABAC
entropy stage cannot be parallelized across samples; the unitary dechirp
transform, $O(\log n)$ FFT-dominated butterflies per sample; and a learned
codec of the class evaluated here, roughly $10^3$--$10^4$ multiply-accumulate
operations per sample, but parallel by construction. By this accounting the
lowest sustaining rate on the Sentinel-1 scene belongs to the most deployable
class in the study, entropy-coded scalar quantization (5.92 bits,
Section~\ref{sec:frontier}), and the registered-grid frontier holder,
raw-domain HEVC at 7.62 bits, is the least deployable; the previous versions
of this paper, which had HEVC alone at the frontier, drew the opposite
conclusion. The reversal is specific to re-compression: the entropy-coding
gain that produces 5.92 is a property of the delivered FDBAQ lattice, and on
unencoded Gotcha data the same gain is 0.1--0.4 bits, which would leave
entropy-coded 4-bit BAQ above raw-domain HEVC there. A learned codec's deployability case rests on convolutional
transforms with a factorized or checkerboard-parallel entropy model, not a
fully autoregressive one, which cannot reach radar-rate throughput in
realistic silicon. This constraint was fixed before any model was trained
(Section~\ref{sec:learned}); its hardware implications are discussed with the
future work in Section~\ref{sec:conclusion}.

\subsection{Learned codec: a bounded negative result}
\label{sec:learned}
A small (0.5M-parameter) two-channel hyperprior codec was trained, with a
factorized-plus-Gaussian-conditional entropy model that is parallel by
construction per the deployability constraint above, on patches drawn from the
invertible focus-transform domain (Section~\ref{sec:data}), on a within-scene
held-out split, under a CPU training budget. The codec minimizes the standard
rate--distortion objective
\begin{equation}
\mathcal{L} = R + \lambda\, D,
\qquad D = \lVert \mathbf{x} - \hat{\mathbf{x}} \rVert_2^2,
\label{eq:rd}
\end{equation}
where $R$ is the entropy-model rate estimate (Section~\ref{sec:limitations}
notes this is not an arithmetic-coded rate) and $\lambda$ trades rate against
distortion; two weightings $\lambda \in \{50, 300\}$ were used. Before
training began we fixed the interpretation rules this section reports under.
If the loss is still materially decreasing at the final training step, the
result is reported as a lower bound, not as a finding about the architecture.
If the trained codec cannot outperform BAQ at matched rate on the two-sided
criterion's own inputs ($P_d$ and spurious count), that is a convergence
failure, not a result. We also registered, before training, a falsifiable
prediction (P2): since squared-error training characteristically blurs rather
than rings, the learned codec's failure mode, if any, was predicted to be the
opposite of JPEG2000's, with controlled spurious-detection counts and $P_d$
eroding as weak scatterers smooth into clutter, rather than a
spurious-detection flood.

Neither run reaches a sustaining rate. One configuration converged to a
degenerate near-zero-rate operating point, a calibration error in which the
loss-weighting parameter was set one to two orders of magnitude too low for
this normalized data. The other failed the pre-declared sanity gate and its
training loss was still descending at the final step. The gate as registered
named BAQ as the matched-rate comparator; no BAQ point exists below 2.1 bits
per complex sample, so raw-domain JPEG2000 at 1.0 bits ($P_d$ 0.390, 4{,}192
spurious) was substituted as the comparator at the learned codec's 0.89-bit
operating point, a surrogate recorded here rather than in the pre-declaration.
Against it the learned codec is worse on both $P_d$ (0.251) and spurious count
(9{,}825). Per the
rules fixed in advance, this is reported as a convergence and calibration
failure and as a lower bound on learned-codec performance on this task, not as
evidence for or against the architecture class. Prediction P2 is neither
confirmed nor refuted: the observed failure signature, low $P_d$ together with
a large spurious count, matches neither predicted pole. The working
hypothesis, unverified and offered as a proposed diagnostic rather than a
finding, is that independent per-tile compression with per-tile RMS
normalization introduces seam discontinuities on the patch grid that focusing
then amplifies into phantom detections: a tiling artifact rather than a
property of the learned transform itself.

The value of this negative result comes from the rules fixed in advance.
Without Section~\ref{sec:prereg}'s discipline, a reader could not distinguish
``the architecture does not work'' from ``the model was undertrained and
miscalibrated on a CPU budget''; this run is the latter.

\subsection{Independent replication on unencoded, government-provided data
(AFRL Gotcha GMTI)}
\label{sec:gotcha}
The results above share one limitation (Section~\ref{sec:data}): Sentinel-1
Level-0 data has already passed through onboard FDBAQ before distribution, so
those results measure re-compression rather than compression of raw ADC
output. That gap is closed here using the AFRL Gotcha GMTI Challenge Problem
dataset~\cite{r17}: motion-compensated, airborne X-band phase history with no
prior onboard compression, publicly released (Public Release \# 88 ABW-09-1031
for the paper, \# 88 ABW-09-0967 for the data) to support algorithm
development on this data type. Unlike the other datasets in this paper it is
also government-provided, and airborne rather than spaceborne.

\begin{figure}[tb]
\centering
\includegraphics[width=\columnwidth]{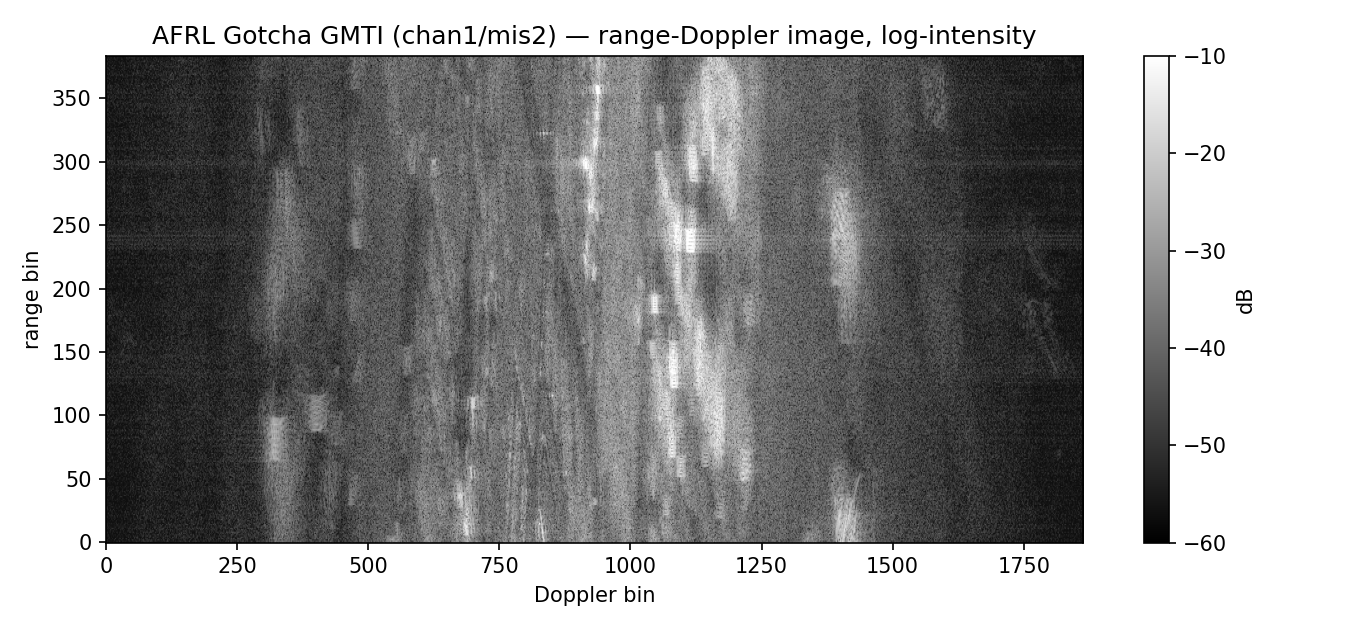}
\caption{The AFRL Gotcha GMTI evaluation scene (channel 1, mission pass,
pulses 5585--7448), the range-Doppler image against which every Gotcha rate,
$P_d$, and spurious-detection number is computed. Formed by the AFRL 2-D IFFT
method and displayed over the $[-60, -10]$~dB range AFRL specifies. Urban
clutter in range-Doppler coordinates: bright near-vertical streaks are strong
stationary scatterers, over which the single ground-truthed moving vehicle
(the Durango) is a small contribution---hence the sparse, 259-peak
reference set.}
\label{fig:gotchascene}
\end{figure}

\textit{Methodology.} The evaluation protocol of Section~\ref{sec:formulation}
is applied with the same CA-CFAR configuration and the same two-sided
criterion, on the raw-domain codec grid (BAQ at 2/3/4/6 bits; JPEG2000 at
ratios 4/8/16/32; HEVC at QP 12/20/28/36). Two grid differences from the
Illinois study are recorded: no low-rate extension sweep was run here, and the
transform-domain HEVC arm used QP 12/20/28/36 rather than Illinois's
36/40/44. The protocol is applied to one channel of one pass (chan1, mis2, pulses 5585--7448, AFRL's own worked
example) of the Durango challenge scene (Figure~\ref{fig:gotchascene}).
Reference and reconstructed images
are formed with AFRL's published range-Doppler method~\cite{r17} rather than
the stripmap RDA focuser used for Sentinel-1, since this data is not
stripmap-geometry raw echo; image formation for motion-compensated phase
history is a direct 2-D IFFT. No parameter of the raw-domain evaluation was
chosen or adjusted after seeing this data. One extension was then added: a
transform-domain arm analogous to the construction of
Section~\ref{sec:data}, testing whether compressing in the focused
(range-Doppler) domain rather than the raw phase-history domain changes the
frontier on this dataset as well. Because Gotcha's phase history is already
motion-compensated, the focus step is a plain 2-D discrete Fourier transform (DFT) pair rather than
Sentinel-1's dechirp/RCMC/azimuth chain; the forward transform is the
image-formation step already used for scoring, the inverse is its exact
adjoint, and the round trip was verified invertible at $4.4\times10^{-7}$ mean
relative error (float32 precision) before use. JPEG2000 and HEVC were each run
in this domain at the same ratio/QP grid.

\begin{figure}[tb]
\centering
\includegraphics[width=\columnwidth]{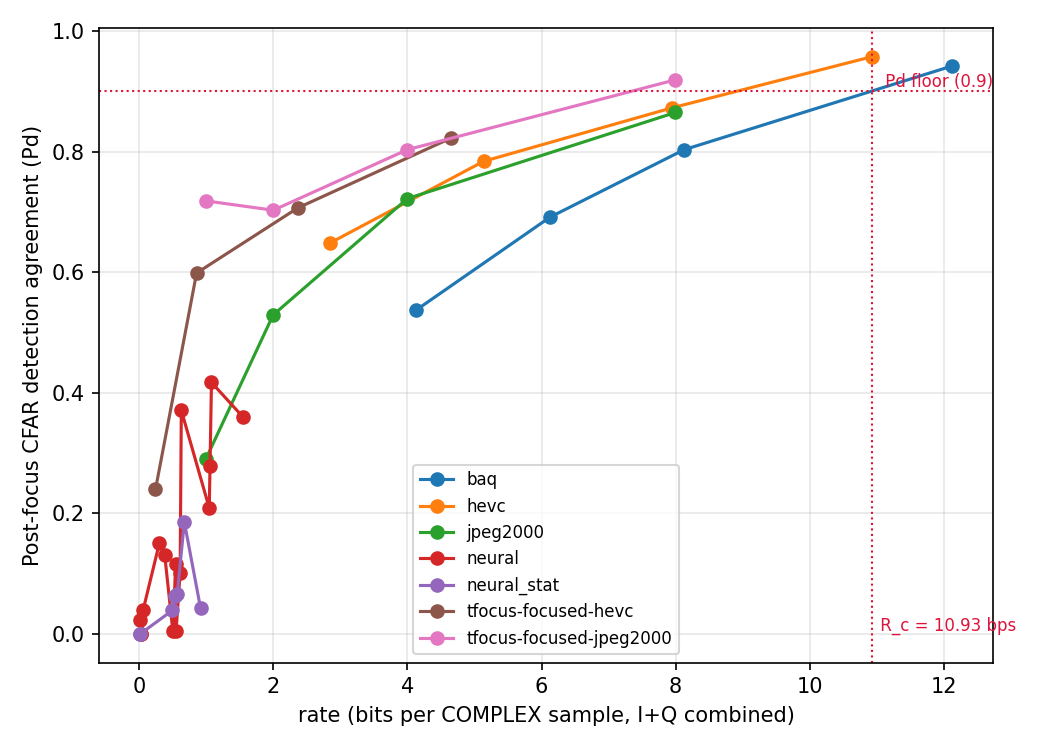}
\caption{Post-focus CFAR detection agreement ($P_d$) vs.\ rate on the Gotcha
GMTI Durango scene (chan1/mis2), raw and focused-domain arms. Horizontal
dotted line: the same $P_d$ floor (0.9) of \eqref{eq:crit} used throughout
this paper. Vertical dotted line: this scene's lowest sustaining rate, 10.93 bits
per complex sample (raw-domain HEVC, QP~12); the frontier is bracketed at
8.0--10.9 (Section~\ref{sec:gotcha}). Corrected scorer.}
\label{fig:pdg}
\end{figure}

\begin{figure}[tb]
\centering
\includegraphics[width=\columnwidth]{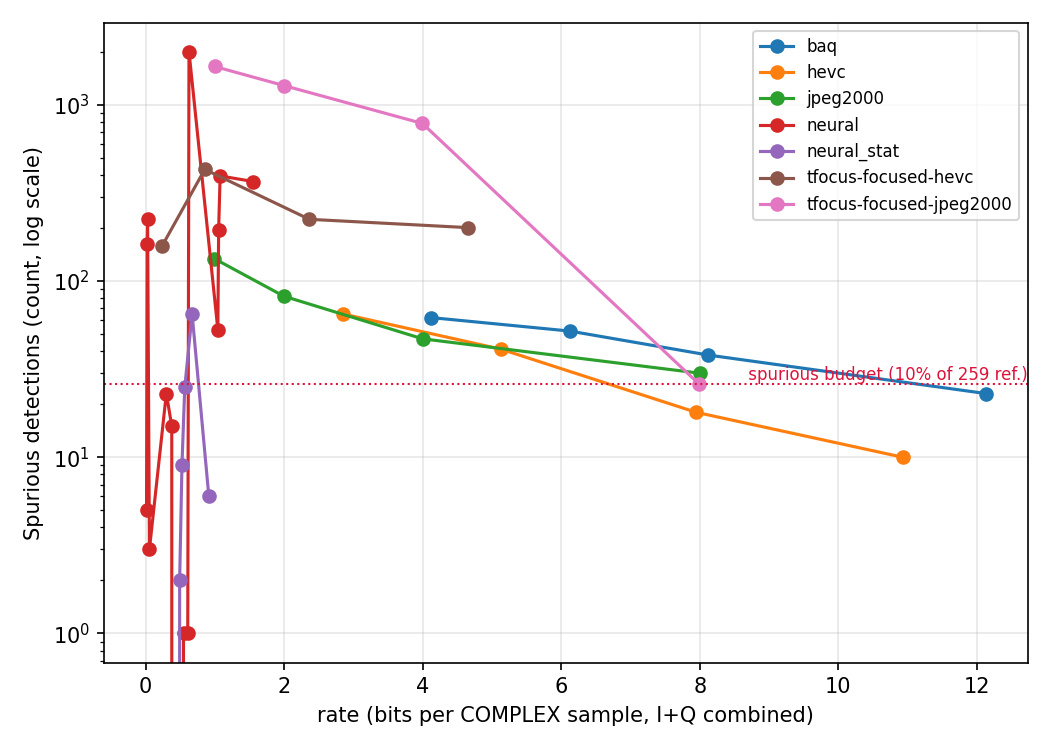}
\caption{Spurious detections vs.\ rate on the same scene, log scale---the
focused-domain arms sit an order of magnitude or more above the raw-domain
codecs at matched low rates, the same spurious-detection signature as
Figure~\ref{fig:spur}. Dotted line: the 25.9-detection budget (10\% of 259
reference detections). Corrected scorer.}
\label{fig:spurg}
\end{figure}

\textit{Results.} The scene yields 259 reference detections (a single dominant
scatterer, the GPS-truthed Durango, plus urban clutter, versus 13{,}964 in the
much larger Illinois crop; Figures~\ref{fig:pdg} and~\ref{fig:spurg}). Under
the two-sided criterion the lowest sustaining rate is 10.93 bits per complex
sample (raw-domain HEVC at QP~12: $P_d$ 0.958, 10 spurious detections against
a budget of 25.9); the next sustaining point is 6-bit BAQ at 12.12. Three
configurations sit just below: focused-domain JPEG2000 at ratio 4 (7.99 bits)
retains 0.919 with 26 spurious, one over budget; raw JPEG2000 at ratio 4 (8.00
bits) retains 0.865; raw HEVC at QP~20 (7.95 bits) retains 0.873. On a
259-detection reference the retention resolution is 0.004 and the budget
resolution is one detection, so the frontier on this scene is reported as a
bracket, $R_c = 8.0$--$10.9$ bits per complex sample, rather than a point.
The focused JPEG2000 row passes the retention floor while exceeding the
spurious budget by a single detection on a 259-peak reference; that is noise
level, and it is not counted as an instance of the certification failure. The
focused-domain rows of this scene are results under the study's plane
mapping, whose no-codec round trip clips 5.4\% of this scene's reference
peaks in that domain (Section~\ref{sec:posthoc}). The
frontier is higher than the Illinois scene's 7.62 bits, for reasons discussed
below, and it is sensitive to the reference's composition: under added noise
the weak references drop out and the frontier falls to about 8 bits
(Section~\ref{sec:posthoc}). Table~\ref{tab:gotcha} lists the operating
points.

\textit{Ground-truth anchor.} The scene's one surveyed target, the Durango,
is stationary during this coherent processing interval (AFRL's GPS record:
speed 0.0~m/s at the three samples spanning it) and sits at AFRL's documented
image position (Doppler pixel 834, range pixel 180), where it is a reference
peak with 29.3~dB contrast over the scene median. Scored post hoc at every
operating point of this section (Section~\ref{sec:posthoc}), it remains a
detection in the same cell at 19 of the 20 points, down to 0.85 bits per
complex sample, and is lost only at focused-domain HEVC QP36 (0.24 bits). The
frontier on this scene is therefore set by the population of weaker reference
detections, not by the strong target; the agreement metric does not lose the
one target whose presence is known independently of the detector.

\begin{table}[tb]
\centering
\caption{AFRL Gotcha GMTI: selected classical operating points under the corrected scorer (peak reduction) and the corrected HEVC pipeline; the full set of 20 rows is in the results log. Rate in bits per complex sample; SQNR is image-domain, dB; the spurious budget is 25.9 (10\% of 259 reference detections). ``Sust.'' = both clauses of \eqref{eq:crit} hold. Focused-domain rows are results under the study's plane mapping, whose no-codec round trip clips 5.4\% of this scene's reference peaks in that domain (Section~\ref{sec:posthoc}).}
\label{tab:gotcha}
\renewcommand{\arraystretch}{1.15}
\setlength{\tabcolsep}{2pt}
\footnotesize
\resizebox{\columnwidth}{!}{%
\begin{tabular}{@{}llrrrrc@{}}
\toprule
Codec & Domain & Rate & SQNR & $P_d$ & Spur. & Sust. \\
\midrule
BAQ (6-bit) & raw & 12.12 & 30.2 & 0.942 & 23 & yes \\
HEVC (QP12) & raw & 10.93 & 34.6 & 0.958 & 10 & yes \\
BAQ (4-bit) & raw & 8.12 & 20.5 & 0.803 & 38 & no \\
JPEG2000 (r4) & raw & 8.00 & 26.5 & 0.865 & 30 & no \\
JPEG2000 (r4) & focused & 7.99 & 20.2 & 0.919 & 26 & no \\
JPEG2000 (r8) & focused & 4.00 & 16.8 & 0.803 & 784 & no \\
HEVC (QP20) & focused & 2.36 & 13.7 & 0.707 & 224 & no \\
JPEG2000 (r16) & focused & 2.00 & 12.2 & 0.703 & 1{,}285 & no \\
JPEG2000 (r32) & focused & 1.00 & 8.8 & 0.718 & 1{,}650 & no \\
HEVC (QP28) & focused & 0.85 & 8.4 & 0.598 & 430 & no \\
Learned (best) & focused & 1.08 & -- & 0.417 & 395 & no \\
\bottomrule
\end{tabular}}
\end{table}

The concentration mechanism of Section~\ref{sec:concentration} replicates on
this dataset. At matched low rates the focused-domain codecs win on $P_d$ (at
about 1 bit per complex sample, focused-domain JPEG2000 holds $P_d$ 0.718
versus raw JPEG2000's 0.290) but produce spurious-detection counts an order
of magnitude above the raw-domain codecs at the same rates (focused JPEG2000:
784--1{,}650 across the low-rate points, versus raw JPEG2000's 47--134;
focused HEVC: 159--430 versus raw HEVC's 41--65). The same trade observed on
Illinois therefore reappears on a different sensor (X-band airborne versus
C-band spaceborne), a different geometry (motion-compensated range-Doppler
versus stripmap RDA), and a different scene (small urban versus large
agricultural), for the wavelet and the block-transform codec alike.

\textit{Interpretation.} The 8.0--10.9 versus 7.62 bits-per-complex-sample
difference is not treated as evidence that classical codecs perform
categorically worse on unencoded data. The two scenes differ in more ways than
encoding history: sensor, resolution, and geometry; a far smaller and sparser
reference-detection set (259 versus 13{,}964, so the per-point $P_d$ and
spurious estimates carry substantially more sampling noise); and a single
dominant point target rather than a distributed agricultural clutter field.
What the result does establish is that the qualitative finding replicates on
independent, government-sourced, unencoded data under a protocol that was never
adjusted for it: a two-sided detection criterion locates a nonzero classical
frontier below fidelity-implied rates, and concentrating energy trades $P_d$
gain for a spurious-detection cost. The specific $R_c$ value is
scene-specific, as Section~\ref{sec:frontier} already states for the Illinois
number; this result supports that scoping.

\textit{Learned codec on this dataset.} The Section~\ref{sec:learned}
hyperprior was also trained on this dataset, with the architecture, lambda
bracket, step budget, and learning rate reused unchanged from the Illinois
configuration. We chose not to retune these for the new dataset, since tuning
until something worked would amount to selecting hyperparameters for the
desired outcome. Training used 13{,}920 focused-domain $64\times64$ patches
drawn from eighty non-overlapping windows tiling the full 154{,}180-pulse
file, with any window overlapping the evaluation crop excluded entirely, and
interpretation rules were again fixed before training. Both runs converged,
with flat final-decile loss slopes, but to a degenerate solution that
transmits nothing: rate collapsed to about 0.02 bits per complex sample within
the first few hundred steps, and reconstruction MSE pinned at exactly 0.500,
which for per-patch RMS-normalized complex data is the variance of the input
itself; the decoder outputs approximately zero. Scored through the identical
pipeline per the rules, both lambdas give $P_d$ 0.003 at 0.02--0.03 bits per
complex sample, and both fail the pre-declared sanity gate. The failure
differs from the Illinois one in an informative way. The Illinois
stronger-lambda run was still descending at its final step (undertrained, a
lower bound); here both runs converged, suggesting that at these loss
weightings, on this data, transmitting nothing is an actual optimum of the
objective as posed. The candidate mechanism, offered as a hypothesis:
per-patch RMS normalization removes cross-patch dynamic range, which is where
the focused domain's energy concentration resides, leaving most patches as
approximately unit-variance complex speckle, a near-memoryless Gaussian source
for which spending no bits is close to rate-distortion optimal. If correct,
this implicates per-tile normalization a second time, independently of
Section~\ref{sec:learned}'s tile-seam hypothesis, from a different failure
signature on a different dataset. It locates the transmit-nothing solution in
the training objective and its normalization rather than in the source,
identifying the collapse as an artifact to be corrected rather than an
intrinsic limit. Whether correcting it suffices to reach a sustaining operating
point, as distinct from merely avoiding the degenerate one, is a separate and
likely harder question.

\textit{Post-draft runs (reported for completeness).} After the first version
of this paper was drafted (2026-08-08), the exploratory phase was reopened on
2026-08-13 and 14 for learned-codec work on the Gotcha data. The 26 runs
appended to the results log in that window are summarized here because the
results log is published as complete; they are post hoc with respect to the
stopping rule of Section~\ref{sec:conclusion}. One run tested the normalization
hypothesis above (notebook Amendment~7): replacing per-patch with global
normalization removed the transmit-nothing collapse (0.63 bits per complex
sample, $P_d$ 0.371, 2{,}004 spurious detections, 77 times the budget),
confirming the hypothesis while failing the sanity gate on spurious count.
Nine runs tested detection-aware and two-sided training losses
(Amendments~8--11; eight under an interpretation rule declared before the
run, the ninth a 100{,}000-step extension of the Amendment~9 run recorded as a
follow-up readout):
a detection-aware loss removed almost every spurious detection at $P_d$
0.10--0.12 (0.54--0.61 bits); a two-sided loss reached $P_d$ 0.21--0.27 with
53--129 spurious at about 1.05 bits after 25{,}000 steps and $P_d$ 0.413 with
621 spurious after 100{,}000 steps; a third loss variant regressed to $P_d$
0.004; an analytic operating-point loss gave $P_d$ 0.13--0.15 with 15--23
spurious at 0.30--0.38 bits. A stage-2 fine-tune and two wider models
(Amendments~12--13) reached $P_d$ 0.28--0.42 with 196--395 spurious at
1.1--1.6 bits. The remaining runs, a detector-in-the-loop variant and a
statistic-domain codec, are recorded in the notebook as readouts without
separately declared interpretation rules and reached at most $P_d$ 0.19. No
configuration met the two-sided criterion or approached the classical
frontier. The best points sit at $P_d$ 0.36--0.42 with 14--24 times the
spurious budget at 1.1--1.6 bits per complex sample (all values under the
corrected scorer; the pixel-count values of the previous version were 0.54--0.59
and 13--21 times), which is the
concentration/spurious-detection trade of Section~\ref{sec:concentration}
reappearing for a learned transform. Further runs in the same window are recorded in the notebook but not in
the results log: GPU fidelity gates on the same image (SQNR only), and two
learned squared-error points scored for detection as well ($P_d$ 0.48 and
0.58 with 1{,}569 and 5{,}081 spurious detections at about 1 bit per complex
sample, under the pixel-count scorer of the previous version; these two were
not re-scored), neither of which met the criterion. They are outside the scope of
this paper. These runs separate the two readings of the negative result only in
part: the collapse was an implementation artifact and is fixed; the residual
trade between retention and spurious detections is unresolved at the
capacities and step budgets tested, and is the open problem stated in
Section~\ref{sec:conclusion}.

\textit{Acknowledgment.} The dataset is credited to AFRL/RYA per its release
conditions (see the Acknowledgment section).

\subsection{Image-domain extension: Umbra spotlight scene (exploratory)}
\label{sec:umbra}
The two scenes above are scored after focusing raw or phase-history data. A
third scene tests the criterion where most operational compression decisions
are made today, on a delivered complex image. The public Umbra Port of Vancouver product (Section~\ref{sec:data}; 22{,}860
reference detections, budget 2{,}286) was compressed directly in the complex
pixel domain with the same plane mapping, codec grid, and CA-CFAR
configuration; there is no focusing step. The evaluation was run on
2026-08-06, after the stopping rule fired, and is exploratory: no endpoint or
threshold was declared for it beyond the criterion already fixed, and nothing
was adjusted after seeing it. Scoring excludes a 128-cell border of the crop
to avoid edge effects. The product's pixels are 32-bit float, so the
compression ratios quoted are against complex64. Table~\ref{tab:umbra} lists
the operating points; the full set of 17 logged rows is in the results log.

\begin{table}[tb]
\centering
\caption{Umbra Port of Vancouver (image domain): selected classical operating points under the corrected scorer (peak reduction) and the corrected HEVC pipeline; the full set of 17 rows is in the results log. Rate in bits per complex sample; SQNR is image-domain, dB; the spurious budget is 2286.0 (10\% of 22{,}860 reference detections). ``Sust.'' = both clauses of \eqref{eq:crit} hold. All rows are results under the study's plane mapping, whose no-codec round trip caps image SQNR at 9.0~dB on this scene (Section~\ref{sec:posthoc}).}
\label{tab:umbra}
\renewcommand{\arraystretch}{1.15}
\setlength{\tabcolsep}{2pt}
\footnotesize
\resizebox{\columnwidth}{!}{%
\begin{tabular}{@{}llrrrrc@{}}
\toprule
Codec & Domain & Rate & SQNR & $P_d$ & Spur. & Sust. \\
\midrule
JPEG2000 (r4) & raw & 8.00 & 8.9 & 0.979 & 591 & yes \\
JPEG2000 (r8) & raw & 4.00 & 8.1 & 0.927 & 3{,}815 & no \\
HEVC (QP12) & raw & 3.87 & 8.6 & 0.930 & 1{,}492 & yes \\
BAQ (1-bit) & raw & 2.12 & 0.9 & 0.018 & 6{,}505 & no \\
JPEG2000 (r16) & raw & 2.00 & 6.8 & 0.910 & 34{,}916 & no \\
JPEG2000 (r64) & raw & 0.50 & 4.8 & 0.829 & 25{,}849 & no \\
\bottomrule
\end{tabular}}
\end{table}

Three observations. First, the classical frontier on this scene is 3.87 bits
per complex sample under the study's plane mapping (HEVC at QP~12: $P_d$
0.930, 1{,}492 spurious), the lowest of the three scenes; without the
mapping's clip HEVC collapses and the frontier is JPEG2000 at ratio 4, 8.00
bits (Section~\ref{sec:posthoc}), while every JPEG2000 verdict is the same
under either mapping. BAQ fails at every tested rate: a scalar quantizer
matched to memoryless raw echoes has nothing to exploit in a concentrated
image and is the wrong tool there, the same conclusion the ATR complement
reaches (Section~\ref{sec:atr}). Second, this scene provides the clear
instances in the corrected record of the failure the two-sided criterion
exists to catch, at two operating points of one codec: JPEG2000 at 4 bits
retains 0.927 of reference detections with 3{,}815 spurious, 1.7 times the
budget, and passes the floor at every matching tolerance tested (not at the
looser false-alarm rate, where it retains 0.874); JPEG2000 at
2 bits retains 0.910, above the floor at the registered tolerance (it falls
below it at 1 and 2 cells), while producing 34{,}916 spurious detections, 15
times the budget. A retention-only rule certifies both at the registered
setting and \eqref{eq:crit} rejects both; both verdicts are reproduced under
the clip-free mapping (Section~\ref{sec:posthoc}). (JPEG2000 at
1 and 0.5 bits, which passed the floor under the pixel-count scorer, now
retain 0.888 and 0.829 and fail both clauses.) The 2-bit point's image-domain SQNR (6.8~dB)
lies below the sustaining ratio-4 point's 8.9~dB, so a fidelity metric orders
them correctly on this scene, yet the same metric places 6-bit BAQ (8.8~dB,
$P_d$ 0.679) above both (Section~\ref{sec:posthoc}). Third, HEVC at 1.74 bits,
with no wavelet stage, retains 0.853 with 7{,}658 spurious detections, 3.3
times the budget: spurious detections follow energy concentration under
transform coding generally.
Because the product's onboard handling before delivery is not disclosed by
the vendor, this scene carries an unknown prior-compression history and is
reported as an image-domain extension, not as a raw-data result.

\section{Post hoc checks added in revision}
\label{sec:posthoc}
The exploratory phase was reopened a second time, on 2026-10-29, for the
checks that peer review of the previous version asked for and that the record
could not answer by text, and a third time the same day for the scorer
correction described in the version note. Each check's endpoint and
interpretation rule was written into the notebook before its first run
(notebook entries ``JSTARS revision'' and ``E0''), with two exceptions
recorded there: the three focused-HEVC sensitivity rows and the first
(full-line) delivered-bitstream measurement were committed with their first
mention, without a prior declaration; the crop-exact bitstream measurement
that replaced the latter was declared before it ran. Every row carries a
\texttt{study} tag and a \texttt{scorer} tag in the results log, and every
number below is from the corrected scorer. In all, 384 post hoc rows were
written under the original scorer and 342 under the corrected one (including
8 noisy-reference rows), plus 30 re-scored learned-codec rows, 8 lattice rows,
3 focused-HEVC sensitivity rows, and 16 control rows; the tables report selected configurations and every row is in the log, and the quantities computed but not tabulated (PSNR, MSE,
phase RMSE, raw-domain SQNR, radial phantom histograms, Durango contrast and
peak offset) are in the log.

\textit{Fidelity metrics against the task.} Image-domain SQNR was computed on
the scored image of every classical configuration and is reported beside
retention and spurious count in Tables~\ref{tab:illinois}, \ref{tab:gotcha},
and~\ref{tab:umbra}. The declared primary endpoint was the Spearman
correlation between SQNR and spurious count among configurations passing the
retention floor; it is $-0.94$ on Illinois ($n = 6$; exact two-sided permutation $p = 0.017$), and the
three and four passing configurations on Gotcha and Umbra are monotone in
SQNR, with the caveat that these sets are nested rate sweeps in which SQNR
adds little beyond rate. The declared secondary endpoint, pairs
of configurations on one scene within 1~dB of SQNR with opposite two-sided
verdicts, is empty on Illinois and Gotcha and finds four pairs on Umbra
(Table~\ref{tab:fidelity}), where the SQNR scale saturates between 7 and 9~dB
across rates from 3.9 to 12.1 bits because the plane mapping's own round trip
caps it at 9.0~dB: 6-bit BAQ (8.8~dB) fails retention while JPEG2000 at
ratio 4 (8.9~dB) sustains, and JPEG2000 at ratio 8 (8.1~dB) fails the budget
while HEVC at QP~12 (8.6~dB) sustains. Two of the four pairs are across
codec families and two within transform coding; the claim that fidelity
metrics pass what the criterion rejects holds on that scene, under that
mapping, and not more.

\begin{table}[tb]
\centering
\caption{Configurations that take part in a within-scene pair with image-domain SQNR within 1~dB and opposite two-sided verdicts (corrected scorer). Four pairs, all on Umbra: two are scalar quantization against transform coding and two are within transform coding; the scene's SQNR scale is capped at 9.0~dB by the plane mapping (Section~\ref{sec:posthoc}).}
\label{tab:fidelity}
\renewcommand{\arraystretch}{1.15}
\setlength{\tabcolsep}{2pt}
\footnotesize
\resizebox{\columnwidth}{!}{%
\begin{tabular}{@{}llrrrrc@{}}
\toprule
Scene & Codec & Rate & SQNR & $P_d$ & Spur. & Sust. \\
\midrule
Umbra Port of Vancouver & BAQ (6-bit), raw & 12.12 & 8.8 & 0.679 & 1{,}222 & no \\
Umbra Port of Vancouver & JPEG2000 (r4), raw & 8.00 & 8.9 & 0.979 & 591 & yes \\
Umbra Port of Vancouver & HEVC (QP12), raw & 3.87 & 8.6 & 0.930 & 1{,}492 & yes \\
Umbra Port of Vancouver & JPEG2000 (r8), raw & 4.00 & 8.1 & 0.927 & 3{,}815 & no \\
\bottomrule
\end{tabular}}
\end{table}

\textit{Matching tolerance and detector $P_{\mathrm{fa}}$.} On the same
reconstructions, detections were re-matched at tolerances of 1, 2, 3, and 5
cells, and re-detected with the reference at design $P_{\mathrm{fa}}$ of
$10^{-3}$, $10^{-4}$, and $10^{-5}$ (Table~\ref{tab:sens}). The declared
rule called a frontier robust only if the sustaining set at the registered
frontier row was unchanged and the frontier moved by at most one grid point.
Under the corrected scorer the Illinois frontier and its six-member
sustaining set are identical at every tolerance and every
$P_{\mathrm{fa}}$, with the reference count ranging from 5{,}662 to 47{,}771;
the Gotcha and Umbra frontiers are identical across tolerances; and
$P_{\mathrm{fa}}$ moves only the Gotcha set (6-bit BAQ leaves at $10^{-3}$;
three configurations enter at $10^{-5}$ on a 99-detection reference) and the
Umbra frontier at $10^{-3}$ (to 8.00). The tolerance and $P_{\mathrm{fa}}$
sensitivity reported for the previous version was an artifact of the
pixel-count scorer.

\begin{table}[tb]
\centering
\caption{Lowest sustaining rate (bits per complex sample) and sustaining-set size [brackets] at each matching tolerance $\tau$ (cells, at $P_{\mathrm{fa}}=10^{-4}$) and design $P_{\mathrm{fa}}$ (at $\tau=3$), corrected scorer. Registered setting: $\tau=3$, $P_{\mathrm{fa}}=10^{-4}$.}
\label{tab:sens}
\renewcommand{\arraystretch}{1.15}
\setlength{\tabcolsep}{2pt}
\footnotesize
\resizebox{\columnwidth}{!}{%
\begin{tabular}{@{}lcccc|ccc@{}}
\toprule
 & $\tau{=}1$ & $\tau{=}2$ & $\tau{=}3$ & $\tau{=}5$ & $10^{-3}$ & $10^{-4}$ & $10^{-5}$ \\
\midrule
Sentinel-1 & 7.62 [6] & 7.62 [6] & 7.62 [6] & 7.62 [6] & 7.62 [6] & 7.62 [6] & 7.62 [6] \\
$n_{\mathrm{ref}}$ & 13{,}964 & 13{,}964 & 13{,}964 & 13{,}964 & 47{,}771 & 13{,}964 & 5{,}662 \\
AFRL Gotcha GMTI & 10.93 [2] & 10.93 [2] & 10.93 [2] & 10.93 [2] & 10.93 [1] & 10.93 [2] & 7.95 [5] \\
$n_{\mathrm{ref}}$ & 259 & 259 & 259 & 259 & 1{,}020 & 259 & 99 \\
Umbra Port of Vancouver & 3.87 [2] & 3.87 [2] & 3.87 [2] & 3.87 [2] & 8.00 [1] & 3.87 [2] & 3.87 [3] \\
$n_{\mathrm{ref}}$ & 22{,}860 & 22{,}860 & 22{,}860 & 22{,}860 & 47{,}140 & 22{,}860 & 14{,}930 \\
\bottomrule
\end{tabular}}
\end{table}

\textit{Signal-to-noise ratio.} Seeded complex white Gaussian noise was
added to the raw input at $-9$, $-6$, $-3$, and 0~dB relative to the signal
power (the native noise floor was not estimated; the levels are added noise,
nothing more), the uncompressed reference was re-formed from the same noisy
input so that the criterion stays at matched threshold on the same scene,
and every classical configuration of Tables~\ref{tab:illinois}
and~\ref{tab:gotcha} was rerun (Table~\ref{tab:snr}). On Illinois the
reference count falls from 13{,}964 to 7{,}322 at 0~dB, the six-member
sustaining set is unchanged at every level, and the frontier drifts from 7.62
to 7.38 bits only because HEVC QP~28's rate falls slightly with noise. On
Gotcha the frontier falls from 10.93 to about 8.0 bits at $-9$ to $-3$~dB and
returns to 10.36 at 0~dB: the noisy reference loses the weak detections that
set the native floor, and the configurations that enter do so at retentions
of 0.903--0.904 on references of 166--217 detections. The Gotcha frontier is
a property of the reference's composition as much as of the codecs, and is
stated at native only.

\begin{table}[tb]
\centering
\caption{Lowest sustaining rate (bits per complex sample) under added complex white noise relative to signal power, corrected scorer; the reference is re-formed from the same noisy input.}
\label{tab:snr}
\renewcommand{\arraystretch}{1.15}
\setlength{\tabcolsep}{2pt}
\footnotesize
\resizebox{\columnwidth}{!}{%
\begin{tabular}{@{}lccccc@{}}
\toprule
Added noise & native & $-9$~dB & $-6$~dB & $-3$~dB & 0~dB \\
\midrule
Sentinel-1 $R_c$ & 7.62 & 7.51 & 7.46 & 7.42 & 7.38 \\
$n_{\mathrm{ref}}$ & 13{,}964 & 12{,}319 & 11{,}209 & 9{,}334 & 7{,}322 \\
AFRL Gotcha GMTI $R_c$ & 10.93 & 8.00 & 8.00 & 7.94 & 10.36 \\
$n_{\mathrm{ref}}$ & 259 & 217 & 188 & 166 & 132 \\
\bottomrule
\end{tabular}}
\end{table}

\textit{Ringing ablation.} In the focused domain of the Illinois scene, four
base codecs were compared: the study's JPEG2000 9/7 wavelet, the reversible
5/3 wavelet at the same byte budget, HEVC over a QP grid, and BAQ as a
non-transform scalar quantizer (which retains almost nothing in this domain,
$P_d \le 0.01$, and so is not an informative control). Phantoms
(reconstruction peaks more than 3 cells from any reference peak) were counted
within 8 cells of three anchor sets of 1{,}397 each: the strongest reference
peaks, the weakest, and random interior cells (Table~\ref{tab:ringing}). The
declared rule kept the statement ``spurious detections concentrate near
strong scatterers'' only if the wavelet's count near strong anchors was at
least twice its count near weak and near random anchors; at 4 bits the ratios
are 1.99 and 2.31, so the statement is withdrawn for the wavelet, and at 2
bits the excess reverses. HEVC's near-strong counts (117--166) exceed its
near-weak and near-random counts by 2 to 13 times, but a second declared
check attributes that excess to the mapping: the no-codec round trip through
the transform domain yields 184 phantoms with 113 near strong anchors and one
near weak anchors, so every transform arm's 110--180 near-strong phantoms are
the mapping's clip of the strongest scatterers, and net of them no codec
concentrates its spurious detections there. Clip-free reruns confirm the
reading from the other side: with a min/max mapping the near-strong excess
vanishes (67/91/80 at QP~16) while HEVC itself collapses (1.75 bits, $P_d$
0.681), because the heavy-tailed image mapped min/max into HEVC's 12-bit
input quantizes the clutter to nothing; the clip is a precondition for HEVC's
focused-domain performance, not only a confound. The focused-domain HEVC grid
also reaches rates the registered sweep (QP~36, 40, 44) never did: at QP~16
and 20 (8.05 and 6.56 bits) it sustains the criterion at the registered
setting, at QP~24 (5.13 bits) it fails the floor at 0.884; the QP~20 row's
robustness is reported in Section~\ref{sec:frontier} (it fails at tolerances
1 and 2 and at $P_{\mathrm{fa}} = 10^{-3}$).

\begin{table}[tb]
\centering
\caption{Focused-domain ablation on Illinois, corrected scorer. Phantoms: reconstruction peaks more than 3 cells from any reference peak. Near: number of phantoms within 8 cells of the 1{,}397 strongest reference peaks, of the 1{,}397 weakest, and of 1{,}397 random interior cells.}
\label{tab:ringing}
\renewcommand{\arraystretch}{1.15}
\setlength{\tabcolsep}{2pt}
\footnotesize
\resizebox{\columnwidth}{!}{%
\begin{tabular}{@{}lrrrrrrc@{}}
\toprule
Base codec & Rate & $P_d$ & Phantoms & Near strong & Near weak & Near random & Sust. \\
\midrule
JPEG2000 9/7 r8 & 4.00 & 0.858 & 5{,}219 & 143 & 72 & 62 & no \\
JPEG2000 9/7 r16 & 2.00 & 0.820 & 16{,}546 & 171 & 255 & 223 & no \\
JPEG2000 5/3 r8 & 4.00 & 0.852 & 5{,}398 & 138 & 72 & 63 & no \\
JPEG2000 5/3 r16 & 2.00 & 0.834 & 21{,}054 & 176 & 325 & 267 & no \\
HEVC QP16 & 8.05 & 0.946 & 804 & 117 & 9 & 10 & yes \\
HEVC QP20 & 6.56 & 0.914 & 1{,}219 & 122 & 10 & 12 & yes \\
HEVC QP24 & 5.13 & 0.884 & 2{,}108 & 118 & 32 & 35 & no \\
HEVC QP28 & 3.66 & 0.822 & 3{,}624 & 142 & 43 & 45 & no \\
HEVC QP32 & 2.17 & 0.720 & 5{,}769 & 166 & 77 & 67 & no \\
BAQ 1-bit & 2.12 & 0.000 & 245 & 0 & 10 & 2 & no \\
BAQ 2-bit & 4.12 & 0.011 & 206 & 2 & 6 & 3 & no \\
\bottomrule
\end{tabular}}
\end{table}

\textit{Plane-mapping control.} The plane mapping of Section~\ref{sec:data} clips each I/Q plane at the 99.99th percentile of $|$value$|$ before any codec, and in a focused image the clipped samples are the strongest scatterers. A no-codec round trip through the mapping (pre-declared control) costs nothing measurable in the raw domain (Illinois: retention 0.9994, one spurious peak; Gotcha: 0.992, none) and little in the transform or image domains (Illinois transform domain: retention 0.995, 184 spurious peaks, 0.7\% of signal energy clipped, image SQNR 21.0~dB; Umbra: retention 0.999, 153 spurious, but 14\% of energy clipped and an image SQNR of only 9.0~dB, which is the ceiling every Umbra fidelity figure sits under; Gotcha image domain: retention 0.985, 5.4\% of reference peaks clipped, which exceeds the 1\% threshold declared for this control, so the Gotcha focused-domain rows are reported as results under this mapping). Focused-domain JPEG2000 at ratio 8 on Illinois rerun with the clip at the 99.999th percentile and with no clip retains 0.858 and 0.864 with 4{,}984 and 4{,}944 spurious peaks against 0.858 and 5{,}219 under the registered mapping: the verdict and the spurious count are unchanged within 5\%, so the mapping stands as a stated, bounded choice for the wavelet on the primary scene. For HEVC it does not: see the ringing paragraph above, and on Umbra the clip-free frontier moves from 3.87 (HEVC) to 8.00 (JPEG2000 at ratio 4) while the JPEG2000 verdicts at ratios 4, 8, and 16 are unchanged (0.976/479 sustains; 0.930/3{,}278 and 0.912/42{,}035 fail the budget).

\textit{Entropy-coded scalar quantization.} The order-0 entropy of the Lloyd--Max BAQ indices, computed from the logged reconstructions with the block scale counted at 8 bits per block per plane (the fixed-rate rows' convention; the notebook pre-declaration said 16 bits, which would add 0.125 bits per complex sample to every entropy-coded row and change no verdict), gives the rate of the same reconstruction under an entropy coder: on Illinois 4.09, 5.34, 5.92, and 8.15 bits per complex sample at 2, 3, 4, and 6 bits (fixed-rate 4.12, 6.12, 8.12, 12.12), on Gotcha 3.99, 5.82, 7.72, and 11.97. The Illinois gain is large because the input is itself FDBAQ output on an 80-level lattice; the Gotcha gain is 0.1--0.4 bits. The Illinois gain does not survive a change of input: under the added noise of the SNR sweep the 4-bit row's index entropy rises from 5.92 to 7.69 bits at the noisiest level (and the lattice adaptation of Section~\ref{sec:harness} moves it to 5.93), so the figure is a property of re-compressing the delivered lattice, and the rates in this paragraph are order-0 bounds, not the output of a coder. The 4-bit Illinois row at 5.92 bits is the lowest sustaining rate of any tested codec on that scene (Section~\ref{sec:frontier}); on Gotcha the entropy-coded 6-bit row (11.97) does not displace raw HEVC at 10.93. The notebook declaration for this check said the entropy-coded row would become the registered-grid frontier holder; the manuscript instead keeps the registered grid's 7.62 and reports 5.92 as the lowest tested rate, for the reason just given, and the departure is recorded in the notebook.

\textit{Impulse response and ground truth.} A unit point target synthesised
at the crop centre by the transform inverse and focused by the range-Doppler
focuser has a $-3$~dB width of 1.0 cell in range and azimuth and peak
sidelobe ratios of $-14.9$~dB (range) and $-13.3$~dB (azimuth), an unweighted
sinc-class response; this characterizes the transform-and-focuser pair's
self-consistency on a synthetic target, not focusing quality on real echoes,
and it is consistent with ringing from the focuser being confined to a few
cells. The Gotcha ground-truth anchor is reported in Section~\ref{sec:gotcha}.
On the Illinois scene, which has no surveyed truth, a proxy subset of the
reference peaks (the top decile by power, 1{,}397 peaks, the same definition as
the ablation's strong scatterers) is scored beside the full set and labelled a
proxy. Every raw-domain BAQ and HEVC configuration retains at least 0.985
of that subset down to 4.1 bits per complex sample and 0.970--0.980 at 2.1 to
3.3 bits (1-bit BAQ: 0.970 against 0.488 of the full set), so the Illinois
frontier, like Gotcha's, is set by the weak majority of reference peaks. Among
raw-domain configurations the subset falls below 0.95 only for JPEG2000 at 2
bits and below (0.797 to 0.415) and for HEVC at QP~44 (0.940); the full-focus
JPEG2000 arm keeps 0.956 of it at 4 bits against 0.858 of the full set. The
declared proxy definition (20~dB over the CFAR training-ring mean) selected
no detections on this scene, because the 2-cell guard ring is inflated by a
scatterer's own sidelobes; the replacement was chosen after that empty result
and is recorded as such.

\textit{Learned codec, re-scored.} The 30 learned-codec rows of
Section~\ref{sec:learned} and the post-draft runs were re-scored from their
archived checkpoints under the corrected scorer. None approaches the
criterion: the best Illinois point retains 0.251 at 0.89 bits (previously
0.397) and the best Gotcha point 0.417 at 1.08 bits (previously 0.593), with
spurious counts that fall with the reference but remain 7 and 15 times the
respective budgets. The negative result is unchanged in kind and stronger in degree.

\section{Limitations}
\label{sec:limitations}
The main results come from a single scene in a single clutter regime
(agricultural/exurban north-central Illinois stripmap).
Section~\ref{sec:gotcha} replicates the qualitative finding on one independent
AFRL Gotcha scene, but that scene is itself a single pass, single channel, and
single ground-truthed vehicle, with a reference-detection count more than an
order of magnitude smaller than the Illinois scene (259 versus 13{,}964); its
result should be read as a lower-confidence, higher-noise data point rather
than as generalization evidence of equal statistical weight, its frontier is
resolved only to a bracket, and its image is a range-Doppler map of a
1{,}864-pulse aperture rather than a focused image. Section~\ref{sec:umbra}
adds one image-domain scene, exploratory and with an undisclosed
prior-compression history. The detection ``task'' is agreement with a
CA-CFAR detector's output on the uncompressed reference, not detection of
surveyed targets: the reference carries the detector's own speckle crossings
(Section~\ref{sec:formulation}), the frontier is set by its weakest members
(Section~\ref{sec:posthoc}), and the one GPS-truthed target
(Section~\ref{sec:gotcha}) is retained far below every frontier. The
classical codec set omits transform-domain BAQ variants (FFT-BAQ,
range-Doppler BAQ), so the frontier is over general-purpose image codecs plus
Lloyd--Max BAQ with and without entropy coding of its indices
(Section~\ref{sec:frontier}); the Level-0 input is already FDBAQ output
delivered at 5.22 bits per complex sample over the crop, below every
sustaining configuration, so the
Illinois result is a re-compression result and the location of a classical
frontier on unencoded spaceborne data remains open
(Section~\ref{sec:harness}). The plane mapping clips each plane at
the 99.99th percentile before any codec; its no-codec round trip costs
0.5\% of retention and 184 spurious detections in the transform domain of
the Illinois scene (113 of them next to the strongest scatterers), clips 5.4\%
of the reference peaks in Gotcha's image domain and 4.3\% in Umbra's, caps
Umbra's image SQNR at 9.0~dB, and is a precondition for HEVC's focused-domain
and image-domain performance (Section~\ref{sec:posthoc}), so every focused-
and image-domain result is a result under that mapping, every near-strong
phantom count is net of it, and on Gotcha and Umbra the mapping is a named
confound. The added-noise levels of Section~\ref{sec:posthoc} are relative to total
signal power because the native noise floor was not estimated; an estimate
from the Level-0 noise packets would convert them to a change in
noise-equivalent sigma zero. Fidelity metrics were not part of the
pre-registered evaluation; computed post hoc, they order spurious counts
within a rate sweep
and separate verdicts only on the image-domain scene, where the mapping caps
the scale. The
matching rule is nearest-neighbour rather than one-to-one and is lenient
toward the flooding failure mode (Section~\ref{sec:formulation}). Broader
cross-scene and cross-clutter generalization, including performance on
regions not represented in training, remains untested at scale. StripMap raw
data over Houston and S\~ao Paulo, matching Asiyabi et al.'s three-city
training set~\cite{r1}, has been acquired and verified but not yet evaluated.
(Their third city, Chicago, has no StripMap coverage in the archive to
acquire.) The train/held-out split for the learned codec is within-scene, by
acquisition chunk, so it measures within-scene generalization only; its rate
is reported as model entropy, not as the output of an actual arithmetic or
range coding pass. CFAR and the frozen CNN classifier are one detector and one
classifier family, not a sweep across detector types. Finally, every
Sentinel-1 result carries the re-compression caveat of Section~\ref{sec:data};
Section~\ref{sec:gotcha} closes that gap for one dataset, though the
controlled comparison (unencoded versus simulated FDBAQ on the same data)
remains open.

\section{Conclusion}
\label{sec:conclusion}
The methodology is the primary contribution of this work: a task-based,
post-focus, two-sided evaluation criterion, whose retention floor was declared
before analysis and whose spurious-detection budget is a declared-before-
analysis amendment (Section~\ref{sec:prereg}), under which no tested
re-quantization of an already-compressed stream is certified below the
stream's own rate (Section~\ref{sec:harness}), that locates the classical
frontier on re-compressed Sentinel-1 data at 7.62 bits per complex sample on
the registered transform-codec grid, robust to matching tolerance and
detector false-alarm rate (same sustaining set) and to added noise (drifting
to 7.38 at the noisiest level, within the declared rule), with entropy-coded scalar
quantization reaching 5.92 as an order-0 bound specific to the delivered
lattice (Section~\ref{sec:posthoc}), that finds the frontier on unencoded
airborne data at 8.0--10.9 bits (Section~\ref{sec:gotcha}), that rejects on an
Umbra image two configurations a retention-only rule certifies
(Section~\ref{sec:umbra}), with a
candidate mechanism for each class of configuration that fails to reach it:
no transform-coding gain in the raw domain, which is why BAQ works; real but
artifact-contaminated gain in the focused domain, which is why transform
coding there does not below about 6 bits per complex sample; and insufficient
concentration from a partial transform, which is why dechirp-only does not.
The open problem these results leave is specific:
capture the coding gain that energy concentration provides without inheriting
the spurious-detection flood that destroys detection. The criterion also suggests a training objective: optimizing a
differentiable surrogate of the two-sided criterion, in place of a
pixel-fidelity loss, would align the quantity a codec is trained on with the
one it is evaluated against. Whether that alignment is sufficient to reach a
sustaining operating point is a separate question that these results do not
settle. A further, practical argument
for the methodology: a criterion declared before analysis, with a regenerable
results record, is the form an
acceptance test must take if learned codecs are ever to be fielded in
operational radar systems, where after-the-fact benchmark claims are not a
sufficient basis for trust.

We adopted an explicit stopping rule for the exploratory phase, fixed before
either the transform baseline or the learned-codec run existed: the phase
would end once the classical baselines and one trained autoencoder had been
scored against the declared criterion, after which further ideas would
become proposed funded work rather than additional unfunded runs. That rule
fired at the point recorded in Section~\ref{sec:learned}. The
Section~\ref{sec:gotcha} replication and the Section~\ref{sec:umbra} scene
were evaluated afterward under the same protocol; extending the evaluation to
new, independently sourced data is distinct from reopening the exploratory
phase, since no criterion, threshold, or codec setting was adjusted for it.
The exploratory phase was reopened after the first draft of this paper, on
2026-08-13 and 14, for the learned-codec runs on the Gotcha data reported at
the end of Section~\ref{sec:gotcha}; the reopening is recorded in the
notebook, the interpretation rules that were declared before individual runs
are identified there, and those results are labelled post hoc here.

The remaining research agenda follows from what the record shows is not yet
known. On the evaluation itself, the post hoc checks of
Section~\ref{sec:posthoc} close the items the previous version listed first
(fidelity metrics, tolerance and $P_{\mathrm{fa}}$ sensitivity, an SNR
sweep, a ringing ablation, the focuser's impulse response, a ground-truth
anchor, and the crop marker); what remains is a tracking-fidelity metric
against the Gotcha GPS record and a second detector family. On the
deployment side, the operation-count accounting of Section~\ref{sec:deploy}
favours convolutional encoders over attention-based architectures, since
attention's quadratic scaling in sequence length and irregular memory access
map poorly onto streaming field-programmable gate array (FPGA) dataflow at
radar sample rates, while convolutional stages stream with local,
fixed-pattern access; this remains a design argument, not a measurement.
Knowledge-distilled, hardware-quantized learned image codecs have been
demonstrated on FPGA fabric~\cite{r16}, which supports the feasibility of
that deployment path without bearing on any raw-radar result here. On
the AFRL data: the controlled comparison of unencoded versus
simulated-FDBAQ versions of the same data, which would isolate whether prior
onboard compression itself shifts $R_c$; a moving-target tracking-fidelity
metric against the dataset's GPS ground truth (position, speed, and heading
over time, in hand but unused), scoring track quality after compression at
each operating point; and extension to the dataset's additional phase centers
and to the range-profile representation. On the tracking metric, one attempt
has already been made and is reported here. Single-channel coherent and
non-coherent change detection between the mis2 and ref4 passes, following
Scarborough et al.'s published method~\cite{r17}, did not produce a defensible
target signature: mean coherence at a candidate location matching the
documented target signature (29.6~dB contrast, against the ingest-verified
29.3~dB) was statistically indistinguishable from the surrounding background
(0.204 versus 0.206), and an initial visual impression of a decorrelated patch
did not survive quantification. Scarborough et al.\ document the same failure
mode for their harder cases, where single-channel change detection is
overwhelmed by clutter and multi-phase-center space-time adaptive processing (STAP) is required. That is the
appropriate next step, a separate build on the dataset's other two phase
centers, and this attempt replaces ``propose to try STAP'' with a specific
reason the simpler method is insufficient here.

\section*{Reproducibility}
The evaluation harness is public under the Apache 2.0
license,\footnote{\url{https://github.com/signalcurrent/radar-codec-eval}}
released on 2026-08-18 with the first version of this paper and updated with
this version. It contains the CFAR detection scoring and the frozen-ATR
pipeline; the invertible dechirp/focus transform with its round-trip
verification test; every classical-baseline codec and sweep configuration
used here (BAQ, JPEG2000, HEVC, and the fairness-mapping code), including the
image-domain evaluation script of Section~\ref{sec:umbra}; the append-only
results log (\texttt{experiments/runs.jsonl}, seed 1337 throughout; the
post hoc and re-scored rows carry \texttt{study} and \texttt{scorer} tags, and
every number in this paper is from rows tagged \texttt{peak-v2}) from which
every classical, learned-codec, ATR, Umbra, and post hoc number in this paper
regenerates, with the exceptions stated below; the table generator
(\texttt{scripts/make\_v4\_tables.py}) that writes every results table from
those rows; the claims ledger (\texttt{reports/claims\_ledger.md}, one line per
quantitative statement with its row) and the review checklist applied
before submission; and the
pre-registration trail as the notebook \texttt{reports/FINDINGS.md} together
with the repository's commit history. The public repository is a filtered
mirror of the author's working repository, produced with
\texttt{git filter-repo} with author and committer dates preserved: the
withheld paths and internal business documents are removed from all history,
programmatic notes in the notebook are redacted with visible markers, and
notebook sections dated 2026-08-14 and 15, concerning GPU fidelity gates on
the Gotcha image and a separate line of work, both outside this paper's
scope, are not included; the notebook's closing note and the README state
what was removed. Commit hashes
cited in this paper (\texttt{08c420d}, \texttt{22df565}) are those of the
working repository, whose row-write and commit timestamps are reproduced in
the mirror's notebook; the mirror's own hashes differ. That chronology is what
makes the declared-before-analysis claim of Section~\ref{sec:prereg}
checkable in its dates, though not, as Section~\ref{sec:prereg} states, in
the unreadness of the rows. Numbers recorded in
the notebook or in test and configuration files rather than in the results
log are: the distinct-value counts of Section~\ref{sec:data}; the round-trip errors of
Sections~\ref{sec:data} and~\ref{sec:gotcha}; the training-set and step
counts of Sections~\ref{sec:learned} and~\ref{sec:gotcha}; the ``still
descending'' loss slope of Section~\ref{sec:learned}; the 1{,}365-chip ATR
test-set size; the time at which results were first read
(Section~\ref{sec:prereg}); and the single-channel change-detection attempt
of Section~\ref{sec:conclusion}. Withheld from the release are the learned
codec's implementation and trained weights, which are proprietary: that
component appears in this paper only as a bounded negative result under
pre-registered interpretation rules (Section~\ref{sec:learned}), and it is
not required to reproduce any classical result or to evaluate a new codec
against the criterion. All data used is public
or openly licensed: Sentinel-1 via Copernicus Open Access; MSTAR public target
chips; and the AFRL Gotcha challenge-problem data via SDMS, approved for public
release with unlimited distribution (Public Release \# 88 ABW-09-0967). No
classified or controlled unclassified information was accessed or used at any
point. Figures~\ref{fig:pd}, \ref{fig:spur}, \ref{fig:atr}, \ref{fig:pdg},
and~\ref{fig:spurg} are generated from the results log by
\texttt{plot\_curves.py}; Figure~\ref{fig:ringing} by
\texttt{make\_ringing\_figure.py}; Figures~\ref{fig:pipeline},
\ref{fig:scene}, and~\ref{fig:gotchascene} are a schematic and two quicklooks.
No figure is hand-edited.

\textit{Version note.} The second version (October 2026) corrected the
characterization of the retention-only counterfactual, stated the commit
chronology of the criterion's two clauses, scoped the frontier to its mapping
rule and the Gotcha frontier to a bracket, added the Umbra scene and the
post-draft learned-codec runs, and demoted the wavelet-ringing attribution to
a hypothesis; no new experiment was run for it. The third version is the
revision prepared for \textit{IEEE Journal of Selected Topics in Applied Earth
Observations and Remote Sensing} after transfer from \textit{IEEE Transactions
on Geoscience and Remote Sensing}. It corrects an implementation error in the
detection scorer that affects every count reported in the first two versions.
The code intended to reduce each CFAR exceedance region to its local maximum
before matching (Section~\ref{sec:formulation}) did not do so: the reduction
it applied is an identity on the exceedance map, so every ``detection'' in
the previous versions was an exceedance pixel, and clustered pixels of one
peak were counted individually. The error was found in a code-against-text
review during preparation of this revision and is recorded in the public
notebook (entry E0). The corrected scorer keeps an exceedance pixel only where
it is the $3\times3$ local maximum of power. All classical configurations were
re-executed deterministically from the same cached inputs under the corrected
scorer, the learned-codec rows were re-scored from their archived
checkpoints, the archived rows remain in the public log under their original
tags, and every corrected row carries \texttt{scorer = peak-v2}. A second
discrepancy was corrected at the same time: HEVC had been run at 12-bit
precision with the I and Q planes coded as a two-frame sequence; it now runs
per plane, all-intra, at an explicitly stated 12 bits, and its rates changed
by up to 0.24 bits per complex sample on Illinois, 0.5 on Umbra, and 0.6 on
Gotcha. The effect on the headline figures
(v2 $\to$ v3): Illinois reference detections 23{,}387 $\to$ 13{,}964 (scorer); Illinois
registered-grid frontier 4.86 (HEVC QP36) $\to$ 7.62 bits per complex sample
(HEVC QP28; scorer, with the HEVC pipeline change moving the QP28 rate from
7.86 to 7.62), and, with entropy coding of the 4-bit BAQ indices and the
delivered FDBAQ bitstream added as rows, the lowest tested sustaining rate is
5.92 (a focused-domain HEVC setting sustains at 6.56 but is not robust)
against a delivered rate of 5.22 over the crop (within this revision the
delivered rate was first stated as the full-line average, 5.28, then measured
over the crop);
3-bit BAQ sustains (0.905) $\to$ fails (0.842); the retention-only certification example on
Illinois (focused JPEG2000 at 4 bits, 0.912 and 2.5 times budget) $\to$ both
clauses reject it (0.858), and the certification failure is now demonstrated
on Umbra (JPEG2000 at 2 bits, 0.910 and 15 times budget); Gotcha reference
369 $\to$ 259 and frontier 7.4--8.0 $\to$ 8.0--10.9;
Umbra reference 41{,}425 $\to$ 22{,}860 (scorer) and frontier 3.38 $\to$ 3.87
(HEVC pipeline change alone); the
Illinois frontier's reported sensitivity to tolerance, false-alarm rate, and
noise $\to$ none. This version also adds the post hoc checks of
Section~\ref{sec:posthoc}, withdraws the ``target-adjacent'' characterization
of the wavelet's spurious detections and, after a further control, the
``concentrated near strong scatterers'' characterization of HEVC's, adds the
delivered-bitstream, memoryless-lossless, and entropy-coded BAQ rows and the
plane-mapping control, and withdraws the claim that the harness brackets the
fielded operating point.
The revised text was checked against the results log and notebook in
separate, AI-assisted adversarial review passes (see Acknowledgment) before
posting.

\section*{Acknowledgment}
Per AFRL's request, the author acknowledges the AFRL Sensors Directorate
(AFRL/RYA) as the source of the Gotcha GMTI dataset of
Section~\ref{sec:gotcha}; results will be shared with the ATR Division as a
matter of courtesy. The Port of Vancouver SICD product of
Section~\ref{sec:umbra} is from Umbra Open Data and is used under CC BY 4.0.
Portions of the manuscript text were drafted with the assistance of an AI
language model (Claude, Anthropic), throughout the manuscript. All technical
content---experimental design, execution, data, analyses, and
conclusions---is the author's own, and all text was reviewed, edited, and
approved by the author. The author declares no conflict of interest.

\end{document}